\documentclass[useAMS,usenatbib,usegraphicx]{mn2e}
\usepackage{amssymb} 
\usepackage{times} 
\usepackage{aas_macros} 
\usepackage{booktabs}
\usepackage{lscape}
\usepackage{ulem}

\title[FM stars II]
{FM stars II: a Fourier view of pulsating binary stars --\\
determining binary orbital parameters photometrically for highly eccentric cases}

\author[Shibahashi, Kurtz \& Murphy] 
{Hiromoto Shibahashi$^1$, Donald W. Kurtz$^2$, and Simon J. Murphy$^{3,4}$\\ 
$^{1}$Department of Astronomy, The University of Tokyo, Tokyo 113-0033, Japan \\
$^{2}$Jeremiah Horrocks Institute, University of Central 
Lancashire, Preston PR1 2HE, UK\\ 
$^{3}$Sydney Institute for Astronomy, School of Physics, The University of Sydney, NSW 2006, Australia\\
$^{4}$Stellar Astrophysics Centre, Department of Physics and Astronomy, Aarhus University, 8000 Aarhus C, Denmark\\
}

\begin{document} 

\maketitle 

\begin{abstract}%
Continuous and precise space-based photometry has made it possible to measure the orbital frequency modulation of pulsating stars in binary systems with extremely high precision over long time spans.
Frequency modulation caused by binary orbital motion manifests itself as a multiplet with equal spacing of the orbital frequency in the Fourier transform.
The amplitudes and phases of the peaks in these multiplets reflect the orbital properties, hence the orbital parameters can be extracted by analysing such precise photometric data alone. We derive analytically the theoretical relations between the multiplet properties and the orbital parameters, and present a method for determining these parameters, including the eccentricity and the argument of periapsis, from a quintuplet or a higher order multiplet. This is achievable with the photometry alone, without spectroscopic radial velocity measurements. We apply this method to {\it Kepler} mission data of KIC\,8264492, KIC\,9651065, and KIC\,10990452, each of which is shown to have an eccentricity exceeding 0.5. Radial velocity curves are also derived from the {\it Kepler} photometric data. We demonstrate that the results are in good agreement with those obtained by another technique based on the analysis of the pulsation phases.
\end{abstract} 


\section{Introduction} 
\label{sec:1}

The study of binary and multiple stars provides the best measurements of stellar parameters, giving direct, nearly model-independent knowledge of stellar masses, 
and radii as well from eclipsing binaries.
These are fundamental to our understanding of stellar structure and evolution, the bedrock upon which much of stellar astrophysics rests. Asteroseismology can now also provide stellar masses and radii, and even ages for individual stars, but our confidence in asteroseismology rests on calibration of asteroseismic masses and radii with those determined from binary stars. 

The primary observational data for binary star studies are the light curve and radial velocity curve. Additional constraints are given by measurement of stellar effective temperature, $T_{\rm  eff}$, from spectroscopy, and, for a few stars, stellar radius from interferometry. In the case of $T_{\rm  eff}$ and interferometric radius, single measurements are sufficient. The light curve and radial velocity curve, however, are by definition time series, and it is these that demand so much time and effort from observers. 

Traditionally, light curves have been measured with ground-based photometry from a single observing site in one or more photometric bandpasses. Radial velocity curves have been measured by obtaining spectra from a single site repeatedly until the orbital phase is well covered. 
So, 
a large amount of observing time is required, often on large telescopes for fainter stars. This is the traditional bottleneck to binary star studies, and until recently the result has been that high precision masses and radii for stars were only known for a matter of hundreds of stars. That all of stellar structure and evolution theory, and beyond that galactic astronomy, should be based on such a sparse base has been a significant concern. 

Now the study of stellar light curves has been revolutionised. Whereas until recently light curves were studied for no more than a few stars from a few sites, for a limited number of nights per year with great gaps in the data, we now have nearly continuous space-based photometry for hundreds of thousands of stars. Whereas from the ground photometric measurements were precise to parts per thousand, or in the best cases, about 10 times better than that, the space data give precision as good as parts per million. The satellite telescope photometry revolution has a pedigree from the Canadian MOST mission\footnote{http://most.astro.ubc.ca/}, through the ESA CoRoT mission\footnote{http://sci.esa.int/corot/} to the NASA {\it Kepler} mission\footnote{http://kepler.nasa.gov}, and it anticipates the upcoming NASA mission TESS\footnote{http://space.mit.edu/TESS/} and ESA mission PLATO\footnote{http://sci.esa.int/plato/}. 

It is the {\it Kepler} mission that has led to the greatest results, with -- as of this writing -- over 4600 exoplanet candidate discoveries\footnote{http://exoplanets.org/}$^{\rm ,}$\footnote{http://exoplanetarchive.ipac.caltech.edu}$^{\rm ,}$\footnote{http://kepler.nasa.gov/Mission/discoveries/}, over 1800 confirmed exoplanets, asteroseismic results for thousands of pulsating stars, and with light curves of more than 2700 eclipsing binary stars\footnote{http://keplerebs.villanova.edu}. The main {\it Kepler} mission data set provides 4\,yr of nearly continuous photometric white light data with 30-min time resolution for 150\,000 stars, with further data for shorter time spans for another 50\,000 stars. The follow-up studies for exoplanets involve many hundreds of astronomers and require large amounts of ground-based telescope time -- often on some of the world's largest telescopes -- for radial velocity measurements necessary to confirm that masses are indeed planetary. The thousands of binary stars also require large amounts of ground-based telescope time to acquire the radial velocities needed for full astrophysical study of the multiple systems. 

The impediment in these studies is the acquisition of the spectra that have traditionally been used to measure radial velocity from the Doppler shift of wavelength of spectral lines. However, Doppler shift can be measured for any astrophysical source of a stable frequency. We have developed two methods to do this directly from {\it Kepler} mission photometric light curves for pulsating stars, {\it without recourse to ground-based observations}, where the pulsation frequencies are the standard, rather than spectroscopic wavelength. These methods obviate the need for any ground-based observations to determine radial velocity, and they provide nearly continuous radial velocity curves that also cover the entire {\it Kepler} 4-yr observational time span -- an unreachable goal with ground-based radial velocity measurements. Our methods are limited to stars with stable pulsation frequencies, but there are thousands 
of these in the {\it Kepler} data with amplitudes high enough to apply our techniques.

Our two methods are the frequency modulation (FM) method \citep{FM2012} and the phase modulation method 
(\citealt{PM2014}; see also \citealt{Balona2014}; \citealt{koen2014}). Our techniques are related to older observed-minus-calculated ($O-C$) methods of studying stellar frequency variability, but the FM and PM techniques both have the advantage that they use the entire data set for maximum signal-to-noise ratio and that they are particularly suited to multi-mode pulsators. The FM technique has the highest possible frequency resolution, and the PM technique is both easily automated and can combine the results from many pulsation frequencies easily. In addition, many studies of pulsating stars for asteroseismic inference begin with the frequency spectrum, and the FM technique shows explicitly the patterns expected in an amplitude spectrum, or power spectrum, for pulsation frequencies undergoing periodic Doppler shifts from binary orbital motion. 

We have demonstrated the validity of the FM method by showing the consistent results obtained from it when compared to a traditional eclipsing binary light curve analysis \citep{kurtzetal2015}. Furthermore, the FM technique is also applicable to non-eclipsing systems and is a powerful tool to determine the orbital parameters.

In practice, the information desired from the radial velocity curve includes the orbital period, the mass function, $f(m_1,m_2,\sin i)$, the orbital semi-major axes, $a_1 \sin i$ and $a_2 \sin i$, the eccentricity, $e$, the angle between the nodal point and the periapsis, $\varpi$, and the time of periapsis passage, $t_{\rm p}$. \citet{FM2012} showed how to derive all of these except $e$ and $\varpi$ from photometric data using the FM method, while \citet{PM2014} showed how to derive all of them from the PM method. In this paper we show how to extract $e$ and $\varpi$ with the FM method. We also show how to derive a radial velocity curve from the FM method, because many investigators have developed software to analyse binary star orbits using light curves combined with radial velocity curves. For example, the pre-eminent program for binary star analysis is now {\sc phoebe}\footnote{http://phoebe-project.org/1.0}, which is designed for radial velocity curve input \citep{Prsa2005}. We emphasise that the FM technique can provide directly all the information that is extracted from the radial velocity curve, but we also provide the method to generate the radial velocity curve itself, both for input to programs such as {\sc phoebe}, and because the radial velocity curve is visually helpful for thinking about the binary star orbit. 

We have discussed Doppler shift to help visualise what our techniques do, but the FM and PM methods do not rely directly on Doppler shift; they are built on the equivalent R\o{}mer time delay. In Sections 2-4 of this paper we derive the relations to determine the mass function, orbital semi-major axis for a pulsating component, eccentricity, $\varpi$  and $t_{\rm p}$, as well as the radial velocity curve, from the frequencies, amplitudes and phases of the components of a multiplet in the Fourier spectrum, which are split by the orbital frequency of the binary star. 

In Section 5, we provide some examples using {\it Kepler} data for stars with high eccentricity. We compare the results with those obtained with the PM method, and discuss the circumstances for which FM or PM is the preferable technique. We show that the method is robust, and that it is capable of detecting companions with masses down to brown dwarf ($m \le 0.08$\,M$_{\odot}$) and even super-Jupiter masses ($m \le 0.03$\,M$_{\odot}$).

\section{The light travel time effect on pulsation in a binary star}
\label{sec:2}

Let us consider a star sinusoidally pulsating with a single angular frequency $\omega_0$ in a binary.
We name the stars `1' and `2', and suppose that star 1 is pulsating.
As the pulsating star moves in its binary orbit, the path length of the light between us and the star varies,  leading to a periodic variation in the arrival time of the signal at Earth. 
The difference in the light arrival time, $\tau(t)$, compared to the case of a signal arriving from the barycentre of the binary system
is given by
\begin{eqnarray}
	\tau (t) := {{1}\over{c}} \int_0^t v_{\rm rad} (t')\, {\rm d}t',
\label{eq:1}
\end{eqnarray}
where $c$ is the speed of the light and $v_{\rm rad}(t)$ denotes the radial velocity, due to the orbital motion, of the star 1 at the time $t$, where the epoch is the time at which the star passes the nodal point (`N' in Fig.\,\ref{fig:01}) directed toward us.
We follow the convention that the sign of the radial velocity is positive when the star is receding from us.
Hereafter we call $\tau (t)$ the ``time delay''. 
With the help of equation (\ref{eq:1}) for the time delay, the observed luminosity variation at time $t$ is written as
\begin{equation}
	\Delta L(t) = A \cos  \left\{ \omega_0\!\left[\, t- \tau(t)\, \right]+ \phi \,\right\},
\label{eq:2}
\end{equation} 
where $A$ is the amplitude of pulsational luminosity variation and $\phi$ denotes the pulsation phase at $t=0$.
In the following we derive the Fourier transform of the luminosity variation with the form of equation (\ref{eq:2}).

\subsection{The radial velocity as a function of time}
\label{sec:2.1}
\begin{figure}
\centering
\includegraphics[width=0.7\linewidth,angle=0]{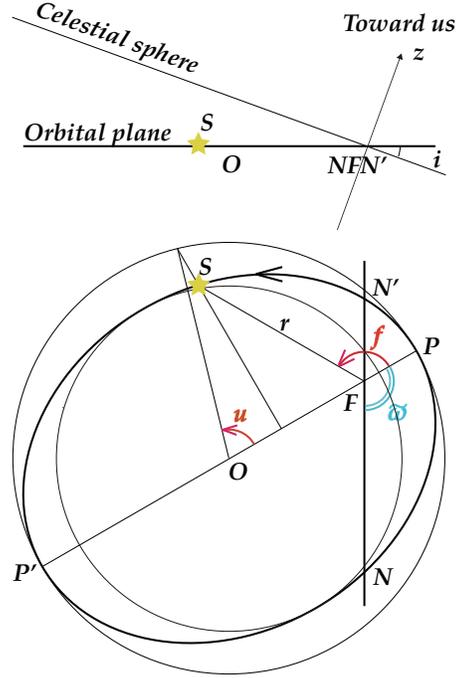} 
\caption{
{\bf Top:} Schematic side view of the orbital plane, seen from a faraway point along the intersection of the orbital plane and the celestial sphere, NFN$^\prime$, where the points N and N$^\prime$ are the nodal points, respectively, and the point F is the barycentre of the binary system; that is, a focus of the orbital ellipses.
The orbital plane is inclined to the celestial sphere by the angle $i$, which ranges from $0$ to $\upi$. In the case of $0 \leq i < \upi/2$, the orbital motion is in the direction of increasing position angle of the star, while in the case of $\upi/2 < i \leq \upi$, the motion is the opposite.
 The $z$-axis is the line-of-sight toward us, and $z=0$ is the plane tangential to the celestial sphere.
{\bf Bottom:} Schematic top view of the orbital plane along the normal to that plane. The periapsis of 
the elliptical orbit is P. The angle measured from the nodal point N, where the motion of the star is directed toward us, 
to the periapsis {\it in the direction of the orbital motion of the star} is denoted as $\varpi$. The star is located, at this moment, at S on the orbital ellipse, for which the focus is F. The semi-major axis is $a_1$ and the eccentricity is 
$e$. Then $\overline{{\rm OF}}$ is $a_1e$. The distance between the focus, F, and the star, S, is $r$. The angle PFS is `the true anomaly', $f$, measured from the periapsis to the star at the moment {\it in the direction of the orbital motion of the star}. `The eccentric anomaly', $u$, also measured {\it in the direction of the orbital motion of the star}, is defined through the circumscribed circle that is concentric with the orbital ellipse.}
\label{fig:01}
\end{figure}

First, we have to derive the radial velocity as a function of time. Let us suppose a plane that is tangential to the celestial sphere on which the barycentre of the binary is located, and let the $z$-axis that is perpendicular to this plane and passing through the barycentre of the binary be along the line-of-sight toward us (see Fig.\,\ref{fig:01}). 
The orbital plane of the binary motion is assumed to be inclined to the celestial sphere by the angle $i$, 
which ranges from $0$ to $\upi$. The orbital motion of the star is in the direction of {\it increasing} position angle, if $0\leq i < \upi/2$, and in the direction of {\it decreasing} position angle, if $\upi/2 < i \leq \upi$.
We write the semi-major axis and the eccentricity of the orbit as $a_1$ and $e$, respectively. 
Let $\varpi$ be the angle measured from the nodal point N, where the motion of the star is directed toward us, 
to the periapsis {\it in the direction of the orbital motion of the star}. Also let $f$ be the `true anomaly', which is the instantaneous angle measured from the periapsis to the star {\it in the direction of the orbital motion of the star},
and let $r$ be the distance between the barycentre and the star\footnote{The range of $i$, the definition of $\varpi$, and the sign of $z$ are often differently given by different authors. It should also be noted here that we cannot distinguish a motion illustrated in Fig.\,\ref{fig:01} and a motion in the inverted image of the bottom panel of Fig.\,\ref{fig:01} from the radial velocity measurement alone. The former is a prograde motion, while the latter is a retrograde motion with the same value of $\varpi$, but with the inclination angle being the supplementary angle of $i$ shown in the top panel of Fig.\,\ref{fig:01}. }. 

To make some complicated formulae derived later in this paper more easily traceable, we must repeat the fundamental derivation of the radial velocity as a function of time, although it was already given in \cite{FM2012}. 
The $z$-coordinate of the position of the star is written as
\begin{equation}
	z = r\sin(f+\varpi)\sin i, 
\label{eq:3}
\end{equation}
and the radial velocity, $v_{\rm rad}:= -{\rm d}z/{\rm d}t$, is 
\begin{equation}
	v_{\rm rad} = -\left[{{{\rm d}r}\over{{\rm d}t}}\sin(f+\varpi) +r{{{\rm d}f}\over{{\rm d}t}}\cos(f+\varpi)\right]\sin i.
\label{eq:4}
\end{equation}
With the help of the known laws of motion in an ellipse 
(e.g., \citealt{brouwerclemence1961}),
\begin{equation}
	r{{{\rm d}f}\over{{\rm d}t}}={{a_1\Omega(1+e\cos f)}\over{\sqrt{1-e^2}}}
\label{eq:5}
\end{equation}
and
\begin{equation}
	{{{\rm d}r}\over{{\rm d}t}}={{a_1\Omega e\sin f}\over{\sqrt{1-e^2}}},
\label{eq:6}
\end{equation}
the radial velocity of star 1 along the line-of-sight is 
expressed as
\begin{eqnarray}
	v_{\rm rad}
	=
	-{{\Omega a_1\sin i}\over{\sqrt{1-e^2}}} 
	\,\left[\cos (f+\varpi) + e\cos\varpi\right],
\label{eq:7}
\end{eqnarray}
where $\Omega$ denotes the orbital angular frequency.
The time dependence of radial velocity
is implicitly expressed by the true anomaly $f$, which can be written
in terms of `the eccentric anomaly', $u$ (see Fig.\,\ref{fig:01}), defined through
the circumscribed circle that is concentric with the orbital ellipse, 
satisfying
\begin{equation}
	r \cos f = a_1(\cos u -e)
\label{eq:8}
\end{equation}
and
\begin{equation}
	r \sin f = a_1\sqrt{1-e^2} \sin u.
\label{eq:9}
\end{equation}
The eccentric anomaly $u$ is written as
\begin{equation}
	u=\Omega (t-t_{\rm p}) + 2\sum_{n=1}^\infty {{1}\over{n}} J_n(ne)\sin n\Omega (t-t_{\rm p}),
\label{eq:10}
\end{equation}
where $J_n(x)$ denotes the Bessel function of the first kind of integer order $n$, and $t_{\rm p}$ denotes the time of the periapsis passage of the star.

The trigonometric functions of the true anomaly $f$ are expressed in terms of a series expansion of trigonometric functions of the time after the star passed the periapsis with Bessel coefficients 
(see Appendix 1):
\begin{equation}
	\cos f = -e + {{2(1-e^2)}\over{e}}\sum_{n=1}^\infty J_n(ne)\cos n\Omega (t-t_{\rm p}),
\label{eq:11}
\end{equation}
\begin{equation}
	\sin f = 2\sqrt{1-e^2} \sum_{n=1}^\infty J_n{'}(ne) \sin n\Omega (t-t_{\rm p}),
\label{eq:12}
\end{equation}
where $J_n{'}(x) := {\rm d}J_n(x)/{\rm d}x$.
The radial velocity is then written explicitly as a function of time:
\begin{equation}
	v_{\rm rad}(t)
	=
	-{\Omega a_1\sin i} 
	\sum_{n=1}^{\infty} n\xi_n(e,\varpi)\cos\,[ n\Omega (t-t_{\rm p}) +\vartheta_n
	],
\label{eq:13}
\end{equation}
where
\begin{eqnarray}
\lefteqn{
	\xi_n(e,\varpi) 
	:=
	2{{\sqrt{1-e^2}}\over{ne}} J_n(ne) 
}
	\nonumber\\
\lefteqn{
\qquad\qquad\quad
	\times
	\sqrt{\cos^2\varpi + \left({{e}\over{\sqrt{1-e^2}}} {{J_n'(ne)}\over{J_n(ne)}}\right)^2 \sin^2\varpi}
}
\label{eq:14}
\end{eqnarray}
and
\begin{equation}
	\vartheta_n(e,\varpi) := 
		\arctan\left[{{e}\over{\sqrt{1-e^2}}} {{J_n'(ne)}\over{J_n(ne)}}\tan\varpi\right] ,
\label{eq:15}
\end{equation}
where $\arctan(x)$ returns the principal value of the inverse tangent of $x$.
\begin{figure}
\begin{center}
	\includegraphics[width=\linewidth, angle=0]{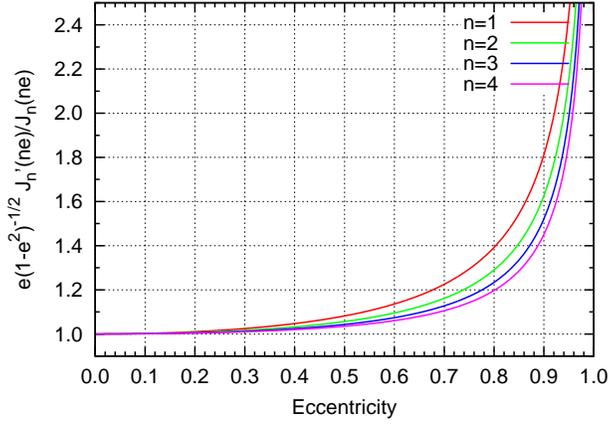} 
\end{center}
\caption{$e (1-e^2)^{-1/2} J_n'(ne)/J_n(ne)$ for $n=1$ (red), $n=2$ (green), $n=3$ (blue), and $n=4$ (magenta) as
functions of eccentricity $e$. This shows that $e (1-e^2)^{-1/2} J_n'(ne)/J_n(ne)$ for $n=1, ..., 4$ is $\sim 1$ except for cases of large $e$, which means $\vartheta(e,\varpi)\simeq \varpi$ except for cases of large $e$.
This characteristic is used later in Sect.\,\ref{sec:3.3}.
}
\label{fig:02}
\end{figure} 

Since 
\begin{eqnarray}
\lefteqn{
	{{J_n'(ne)}\over{J_n(ne)}} 
	=
	{{1}\over{e}}
	-
	\left[
	{{ne/2}\over{(n+1)}} \mathop{}_{-} {{n^2e^2/2}\over{(n+2)}} \mathop{}_{-} 
	{{n^3e^3/2}\over{(n+3)}} \mathop{}_{-} \dots \right] ,
}
\label{eq:17}
\end{eqnarray}
where the series in square brackets denotes continued fractions, the $\varpi$-dependence of $\xi_n(e,\varpi)$ is weak for $e \ll 1$,
but it becomes substantial with the increase of $e$ (see Fig.\,\ref{fig:03}).
Also, since 
\begin{eqnarray}
	J_n(x)=\sum_{k=0}^\infty (-1)^k {{(x/2)^{n+2k}}\over{(n+k)!\,k!}},
\label{eq:18}
\end{eqnarray}
the lowest order of the series expansion of $\xi_n(e,\varpi)$ in terms of power of $e$ is $e^{n-1}$; 
$\xi_n(e,\varpi) \sim O(e^{n-1})$ for $e\ll 1$.

\subsection{The time delay as a function of time}
\label{sec:2.2}
Integrating equation (\ref{eq:13}), we obtain an expression for the time delay as a function of time
\begin{equation}
	\tau (t) 
	=
	-{{a_1\sin i}\over{c}}
	\sum_{n=1}^\infty \xi_n
	\sin [n\Omega (t-t_{\rm p}) + \vartheta_n]
	-\tau_0 ,
\label{eq:19}
\end{equation}
where
\begin{eqnarray}
	\tau_0(e,\varpi) := -{{a_1\sin i}\over{c}}\sum_{n=1}^\infty \xi_n\sin (-n\Omega t_{\rm p}+\vartheta_n) .
\label{eq:20}
\end{eqnarray}

\begin{figure}
\begin{center}
	\includegraphics[width=\linewidth, angle=0]{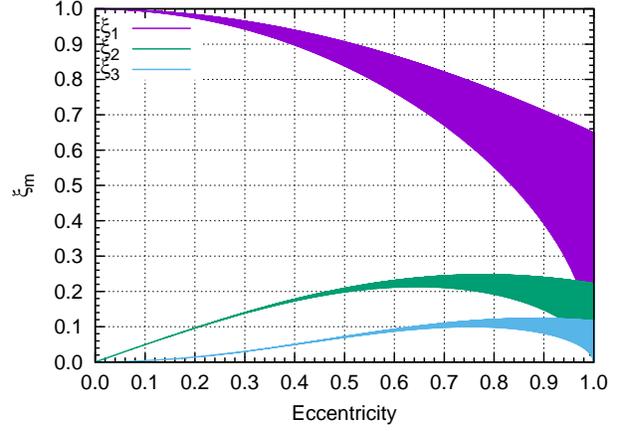} 
\end{center}
\caption{The coefficients of $\xi_1(e, \varpi)$ (purple), $\xi_2(e, \varpi)$ (green), $\xi_3(e, \varpi)$ (blue), 
and $\xi_4(e, \varpi)$ (magenta) for the Fourier amplitudes of the phase modulation as
functions of eccentricity $e$. The band of each curve shows the range of $\varpi$ from $0$ to $2\upi$.
The $\varpi$-dependence of $\xi_2$, $\xi_3$, and $\xi_4$ is substantially weaker than $\xi_1$ in a wide range of $e$.
This figure is used later in Sect.\,\ref{sec:3.3}.
}
\label{fig:03}
\end{figure} 

Fig.\,\ref{fig:03} shows $\xi_1$, $\xi_2$, $\xi_3$, and $\xi_4(e,\varpi)$ as functions of eccentricity $e$. The band of each curve shows the range of $\varpi$ from $0$ to $2\upi$. It is seen that the $\varpi$-dependence of $\xi_2$, $\xi_3$, and $\xi_4$ is substantially weaker in a wide range of $e$ than is the case of $\xi_1$. This is because the terms $\xi_2$, $\xi_3$, $\xi_4$ and the higher terms represent a departure from a pure sinusoidal variation with the orbital angular frequency $\Omega$, which is essentially caused by the eccentricity of the orbit.
On the other hand, $\xi_1(e,\varpi)$ shows substantial $\varpi$-dependence, in particular with increase in $e$. This is because the projected light travel time is dependent not only on the eccentricity of the orbit, but also on the direction of the apparent elongation of the projected orbit, particularly in cases of high eccentricity.

\subsection{Phase modulation}

\cite{FM2012} define a parameter $\alpha$ that measures the amplitude of the phase modulation when the pulsation frequency is treated as fixed:
\begin{equation}
	\alpha := {{a_1\omega_0\sin i}\over{c}} ,
\label{eq:21}
\end{equation}
which is the ratio between the light travel time across the projected semi-major axis and the pulsation period of the mode in consideration. It should be noted that the value of $\alpha$ is dependent on the mode frequency. The larger the orbit size and the shorter the pulsation period, the larger the resulting phase modulation. Then the observed luminosity variation, whose amplitude is assumed to be unity, is rewritten as 
\begin{eqnarray}
	\Delta L (t) 
	= 
	\cos\left[ (\omega_0 t + \varphi)
	+ \alpha\sum_{n=1}^{\infty} \xi_n \sin\left(n\Omega t + \theta_n \right) \right] ,
\label{eq:22}
\end{eqnarray}
where
\begin{equation}
	\varphi := \phi + \omega_0\tau_0
\label{eq:23}
\end{equation}
and
\begin{equation}
	\theta_n := \vartheta_n -n\Omega t_{\rm p}.
\label{eq:24}
\end{equation}
If we regard equation (\ref{eq:22}) as luminosity variation with the intrinsic angular frequency $\omega_0$, the pulsation phase is regarded as being modulated by the orbital motion.

\section{Methodology for determining binary parameters from photometry}  
\label{sec:3}
The problem we address here is how to reproduce the radial velocity $v_{\rm rad}(t)$ from the observed luminosity variation given by equation (\ref{eq:22}), where $v_{\rm rad}(t)$ is rewritten as 
\begin{equation}
\lefteqn{
	v_{\rm rad}(t) 
	=
	-{{\Omega c}\over{\omega_0}} \alpha \sum_{n=1}^{\infty} n\xi_n \cos[n\Omega t +\theta_n] ,
}
\label{eq:25}
\end{equation}
or equivalently, how to deduce the orbital parameters, that is, the orbital angular frequency $\Omega$, the semi-major axis $a_1$, the eccentricity $e$, the angle between the periapsis and the nodal point $\varpi$, 
and the time of periastron passage of the star $t_{\rm p}$. 

\subsection{Fourier transform of the luminosity variation}
\label{sec:3.1}
Here we describe the Fourier analysis of the luminosity variation, given by equation (\ref{eq:22}), showing periodic phase modulation. This is done with the help of Bessel functions, written with arbitrary real $x$ and $\theta$:
\begin{equation}
	{\rm e}^{\pm {\rm i}x\sin\theta}=\sum_{n=-\infty}^{\infty} J_n(x){\rm e}^{\pm {\rm i}n\theta}.
\label{eq:26}
\end{equation}
Applying this, we rewrite equation (\ref{eq:22}) as 
\begin{eqnarray}
\lefteqn{
\Delta L(t) =
	\Re \left[ {\rm e}^{ {\rm i}(\omega_0t+\varphi)}
	\prod_{n=1}^{\infty} {\rm e}^{ {\rm i}\alpha\xi_n\sin\left(n\Omega t+\theta_n\right)} \right]	
}
	\nonumber\\
\lefteqn{
	\quad
	=
	\Re  \left\{ {\rm e}^{ {\rm i}(\omega_0t+\varphi)}
	\lim_{N\rightarrow\infty} \sum_{k_1=-\infty}^\infty \cdots \sum_{k_N=-\infty}^\infty 
	\left[
	\prod_{n=1}^{N} J_{k_n}(\alpha \xi_n)
	\right]
	\right.
}
\nonumber\\
\lefteqn{
\qquad
	\left.
	\times \exp \left[ {\rm i} \sum_{n=1}^N k_n(n\Omega t + \theta_n) \right]
	\right\},
}	
\label{eq:27}
\end{eqnarray}
where $\Re(x)$ means the real part of $x$ and $N$ denotes a large number.
This means that a frequency multiplet appears around the intrinsic frequency $\omega_0$ in the frequency spectrum, and that each component of the multiplet is separated from its neighbouring components by the orbital frequency $\Omega$:
\begin{eqnarray}
\lefteqn{
\quad
\Delta L(t) =
	\Re \biggl\{ {\cal A}_0 {\rm e}^{{\rm i}\omega_0 t} 
}
\nonumber\\
\lefteqn{
	\quad\quad\qquad
	\left.
	+ \sum_{m=1}^{\infty} 
	\left( {\cal A}_{+m} {\rm e}^{{\rm i}(\omega_0 + m\Omega)t}
	+  {\cal A}_{-m} {\rm e}^{{\rm i}(\omega_0 - m\Omega)t} \right) 
	\right\} ,
}
\label{eq:28}
\end{eqnarray}
where ${\cal A}_0$ and ${\cal A}_{\pm m}$ are the complex amplitudes of the central peak and the sidelobes separated from the central component by $\pm m\Omega$, respectively.

\subsection{Truncation in the case of $\mbox{\boldmath$\alpha <1$}$}
\label{sec:3.2}
It is straightforward to compute numerically the terms in equation (\ref{eq:27}) for a given set of binary parameters. The purpose of the present paper is, however, the reverse: we deduce the binary parameters from the Fourier transform of the light curve.
An important point is that the complex amplitudes, ${\cal A}_0$ and ${\cal A}_{\pm m}$, are determined by the quantities that characterize the binary system, $\alpha$, $\xi_n(e,\varpi)$, $\theta_n(e, \varpi)$, allowing us to inversely determine those binary parameters from the amplitude spectrum of the oscillations. 

In the following, we derive analytical expressions of the Fourier transform of the light curve in terms of the binary parameters. In practice, we truncate the infinite expansion series in  equation (\ref{eq:27}) with a finite number of terms, $N$. There appear $(2N+1)^N$ cross-terms, and it is neither practical nor necessary to take account of all the terms. Rather, it is better to keep only the leading terms. Indeed, in the case of $\alpha < 1$, the Bessel functions with the argument of $\alpha\xi_n$ become negligibly small with the increase of $n$ and the order $k_n$. Hereafter, we limit ourselves to consider the cases of $\alpha < 1$; that is, the cases for which the light travel time across the projected orbit is longer than the pulsation period.  

Later, in Section 7, we will demonstrate some observational examples that show up to septuplets, nonuplets, and even undecuplets. So, here we truncate the infinite series of equation (\ref{eq:27}) with $N=5$. Then, equation (\ref{eq:27}) produces $(2N+1)^N=11^5=161051$ cross-terms. Among them, we discard terms separated from the central peak at $\omega_0$ further than $\pm 6\Omega$ and keep only the undecuplet. For each frequency component of $m\Omega$, we keep only the terms up to ${\cal O}(\alpha^2)$.

Then, equation (\ref{eq:28}) shows that the light-time effect in a binary star leads to a frequency undecuplet, 
$\omega=\omega_0 \pm m\Omega$ $(m=1, ..., 5)$,
for which complex amplitudes of the first and second sidelobes are given as
\begin{eqnarray}
\lefteqn{
	{{{\cal A}_{\pm 1}}\over{{\cal A}_0}}
	=
	\pm \left({{ J_{1}(\alpha\xi_1) }
	\over
	{
	 J_0(\alpha\xi_1)  }}
	 \right)
	 \,{\rm e}^{\pm {\rm i}\theta_1}
}
	\nonumber\\
\lefteqn{
	\quad\quad\qquad
	\times \left[
	1\mp 
	\left({{ J_{1}(\alpha\xi_2) }
	\over
	{ J_0(\alpha\xi_2) }}\right)
	\,{\rm e}^{\pm {\rm i}(\theta_2 - 2\theta_1)}
	\right]
}
\label{eq:29}
\end{eqnarray}
and
\begin{eqnarray}
\lefteqn{
	{{{\cal A}_{\pm 2}}\over{{\cal A}_{0}}} 
	\simeq 
	\pm \left( {{J_{1}(\alpha\xi_2)}\over{J_{0}(\alpha\xi_2)}} \right) \,{\rm e}^{\pm {\rm i} \theta_2}
}
	\nonumber\\
\lefteqn{
	\quad\quad\qquad
	\times 
	\left[1\pm \left( {{J_{1}(\alpha\xi_2)}\over{J_{0}(\alpha\xi_2)}} \right)^{-1}
	\left( {{J_{2}(\alpha\xi_1)}\over{J_{0}(\alpha\xi_1)}} \right) 
	\,{\rm e}^{\mp {\rm i}(\theta_2-2\theta_1)} 
\right],
}
\label{eq:30}
\end{eqnarray}
respectively.
The complex amplitudes of the higher order sidelobes are given in Appendix 2.

\subsection{Amplitude ratios and phase differences between the sidelobes and the central component}
\label{sec:3.3}
Let $A_0$ and $\phi_0$ be the amplitude and the phase of the central component of the multiplet at the angular frequency $\omega_0$, 
and let $A_{\pm m}$ and $\phi_{\pm m}$ be the amplitudes and the phases of the $\pm m$-th sidelobes, respectively. 
In the following, we derive the amplitude ratios and the phase differences between the sidelobes and the central component.

\subsubsection{The case of the first sidelobes}
\label{sec:3.3.1}
From equation (\ref{eq:29}), the amplitude ratio between the first sidelobes ($m=1$) and the central peak is given,
up to the order of ${\cal O}(\alpha\xi_1)$,  by 
\begin{equation}
	{{A_{+1}+A_{-1}}\over{A_0}} 
	\simeq
	2\left({{  J_{1}(\alpha\xi_1) }
	\over
	{ J_0(\alpha\xi_1)  }} \right)
\label{eq:31}
\end{equation}
and
\begin{equation}
	{{A_{+1}-A_{-1}}\over{A_{+1}+A_{-1}}}
	\simeq
	-
	\left(
	{{ J_{1}(\alpha\xi_2) }
	\over
	{ J_0(\alpha\xi_2)  }}
	\right)
	\cos(2\vartheta_1 -\vartheta_2)	.	
\label{eq:32}
\end{equation}
Equation (\ref{eq:32}) indicates whether $A_{+1}$ or $A_{-1}$ has the higher amplitude, according to the sign of $\cos (2\vartheta_1-\vartheta_2)$. 
If $e \ll 1$, $\vartheta_i \simeq \varpi$ (see Sect.\,\ref{sec:2.1}). Hence, in such a case, whether $A_{+1}$ or $A_{-1}$ has the higher amplitude is determined by $\cos\varpi$. For $0 \leq \varpi < \upi/2$ or $3\upi/2 < \varpi \leq 2\upi$, $A_{-1} > A_{+1}$,
while, for $\upi/2 < \varpi < 3\upi/2$, $A_{+1} > A_{-1}$. When $\varpi=\upi/2$ or $\varpi=3\upi/2$, that is, when the periapsis or the apoapsis is towards the observer, the amplitudes of $A_{+1}$ and $A_{-1}$ are equal to each other.

Also, the phase differences between the first sidelobes and the central component are given as
\begin{eqnarray}
\lefteqn{
	 \phi_{\pm 1} -\phi_{0} := {\rm arg}\left({{{\cal A}_{\pm 1}}\over{{\cal A}_0}}\right) \pm \Omega t
}
	\nonumber\\
\lefteqn{
	\quad\qquad
	\simeq
	\left(
	{{\upi}\over{2}} \mp {{\upi}\over{2}}
	\right)
}
	\nonumber\\
\lefteqn{
\quad\quad\qquad
	\pm \left(
	\Omega t 
	+ \theta_1 
	\right)
	+ \left( {{J_1(\alpha\xi_2)}\over{J_0(\alpha\xi_2)}} \right)
	 \sin(2\vartheta_1-\vartheta_2).
}	 
\label{eq:33}
\end{eqnarray}

Let us choose the zero point for the phases so that the phases of the first sidelobes are equal. This condition is fulfilled twice during one orbital period, at $t=t_{0}$ being
\begin{equation}
\lefteqn{
	\Omega t_{0} + \theta_1 = \left(k+{{1}\over{2}}\right)\upi,
}
\label{eq:34}
\end{equation}
where 
\begin{equation}
\lefteqn{
	k=
	\left\{ \begin{array}{llll}
	0  &\mbox{and}\ & 1 & \mbox{if}\ 0\leq \theta_1\leq \upi/2 \\
	1  &\mbox{and}\ & 2 & \mbox{if}\ \upi/2 \leq \theta_1 \leq 3\upi/2 \\
	2  &\mbox{and}\ & 3 & \mbox{if}\ 3\upi/2 \leq \theta_1 < 2\upi .
	\end{array}\right.
}
\label{eq:35}
\end{equation}
At each of these phases, the Fourier component proportional to $\cos\Omega t$ of the radial velocity becomes zero.
In the case of a circular orbit, this corresponds to the moment at which the orbital motion of the star is perpendicular to the line-of-sight. 
At $t=t_{0}$,
\begin{eqnarray}
\lefteqn{
	\phi_{\pm 1}(t_{0})-\phi_0
	=
	\pm \left( k \pm {{1}\over{2}} \right) \upi 
}
	\nonumber\\
\lefteqn{
\quad\qquad\qquad\qquad
	+\left( {{J_1(\alpha\xi_2)}\over{J_0(\alpha\xi_2)}} \right) \sin(2\vartheta_1-\vartheta_2),
}
\label{eq:36}
\end{eqnarray}
which means that {the phases of the first sidelobes apparently differ from that of the central peak approximately by either $+\upi/2$ or $-\upi/2$}. This can be used to confirm that FM is caused by the orbital motion. Equation (\ref{eq:36}) also indicates that the deviation of the phase difference from ${+\upi}/2$ or $-\upi/2$ is dependent on the sign of $\sin(2\vartheta_2-\vartheta_1)$. Since $\vartheta_i \simeq \varpi$, this means that {the phase difference is slightly larger (smaller) if the periapsis is in the far (near) side of the orbit, with respect to us}.

\subsubsection{The case of the second sidelobes}
\label{sec:3.3.2}
Similarly, up to the order of ${\cal O}(\alpha\xi_2)$,
\begin{eqnarray}
\lefteqn{
	{{A_{+2} + A_{-2}}\over{A_0}} 
	\simeq
	2
	\left( {{J_1(\alpha\xi_2)}\over{J_0(\alpha\xi_2)}}\right)
}
\label{eq:37}
\end{eqnarray}
and
\begin{equation}
\lefteqn{
	{{A_{+2} - A_{-2}}\over{A_{+2}+A_{-2}}} 
	\simeq
	\left( {{J_{1}(\alpha\xi_2)}\over{J_{0}(\alpha\xi_2)}} \right)^{-1}
	\left( {{J_2(\alpha\xi_1)}\over{J_0(\alpha\xi_1)}} \right)
	\cos (2\vartheta_1 - \vartheta_2).	
}
\label{eq:38}
\end{equation}
Equation (\ref{eq:38}) indicates whether $A_{+2}$ or $A_{-2}$ has the higher amplitude is determined by the sign of $\cos (2\vartheta_1-\vartheta_2)$. Note that if $A_{+1}$ is higher (lower) than $A_{-1}$, then $A_{+2}$ is lower (higher) than $A_{-2}$.

Also the phase differences are 
\begin{eqnarray}
\lefteqn{
	 \phi_{\pm 2} -\phi_{0} := {\rm arg}\left({{{\cal A}_{\pm 2}}\over{{\cal A}_0}}\right) \pm 2\Omega t
}
	\nonumber\\
\lefteqn{
	\quad\qquad
	\simeq
	\left(
	{{\upi}\over{2}} \mp {{\upi}\over{2}}
	\right)
	\pm \left(
	2\Omega t 
	+ \theta_2 
	\right)
}
	\nonumber\\
\lefteqn{
\quad\quad\qquad
	- \left( {{J_{1}(\alpha\xi_2)}\over{J_{0}(\alpha\xi_2)}} \right)^{-1}
	\left( {{J_2(\alpha\xi_1)}\over{J_0(\alpha\xi_1)}} \right)
	 \sin(2\vartheta_1-\vartheta_2).
}	 
\label{eq:39}
\end{eqnarray}
Hence, at $t=t_{0}$, 
\begin{eqnarray}
\lefteqn{
	 \phi_{+2} (t_{0}) -\phi_{-2} (t_{0})
	 = 
	 -\upi + 2(2\Omega t_{0} + \theta_2).
}
\label{eq:40}
\end{eqnarray}

Combining equations 
(\ref{eq:34}) and (\ref{eq:40}), we obtain 
\begin{eqnarray}
\lefteqn{
	 2(\theta_2 - 2\theta_1 )
	 =
	 \left[\phi_{+2} (t_{0}) -\phi_{-2} (t_{0})\right] -(4k+1)\upi ,
}
\label{eq:41}
\end{eqnarray}
equivalently,
\begin{equation}
	2\vartheta_1 - \vartheta_2 
	=
	- {{1}\over{2}} \left[\phi_{+2} (t_{0}) -\phi_{-2} (t_{0})\right] +{{\upi}\over{2}}.
\label{eq:42}
\end{equation}
It should be noted that there remains an uncertainty of $2n\upi$ in the measured phase difference, $\phi_{+2}(t_{0})-\phi_{-2}(t_{0})$, where $n$ is an integer, and then we cannot uniquely determine $(2\vartheta_1 - \vartheta_2)$ from this equation alone. This uncertainty can be eliminated, however, by taking account of the inequality of $A_{\pm 1}$ and the deviation of $\phi_{\pm 1}(t_{0})-\phi_0$ from $\pm \upi/2$.

\subsubsection{General description: the higher-order sidelobes} 
\label{sec:3.3.3}
Similarly, up to ${\cal O}(\alpha\xi_{m})$, for the $m$-th sidelobes ($m \geq 1$), 
\begin{equation}
\lefteqn{
	{{A_{+m} + A_{-m}}\over{A_0}} 
	\simeq
	2
	\left( {{J_1(\alpha\xi_m)}\over{J_0(\alpha\xi_m)}}\right).
}
\label{eq:43}
\end{equation}
Since the value of the left-hand side is observationally obtained, this equation determines the value of $\alpha\xi_m$. 
A practical numerical method of solving equation (\ref{eq:43}) is given in Section\,\ref{sec:4.1}. 
The graphic solutions are shown in Fig.\,\ref{{fig:04}}. 
Also,
\begin{equation}
\lefteqn{
	\phi_{+m}(t_{0})-\phi_{-m}(t_{0}) 
	= 
	-\upi + 2(m\Omega t_{0} + \theta_m).
}
\label{eq:44}
\end{equation}
This relation allows us to deduce
\begin{equation}
\lefteqn{
	2(\theta_m - m\theta_1)
	=
	\left[{\phi_{+m}(t_{0})-\phi_{-m}(t_{0})}\right] 
	-(m-1){{\upi}}.
}
\label{eq:45}
\end{equation}

\subsection{Formal reconstruction of radial velocity from photometry}

Equation (\ref{eq:43}) indicates that the value of $\alpha\xi_m$ is 
observationally determined from the amplitude ratio between the sum of the sidelobes and the central peak.
The graphical solution is shown in Fig.\,\ref{{fig:04}}.
The solution is numerically obtained with sufficient accuracy by expanding the Bessel functions $J_n(x)$ up to the fifth power of $x$.
On the other hand, the phase differences between the sidelobes determine $\theta_m$ from equation (\ref{eq:45}), where $\theta_1$ is given by equation (\ref{eq:34}). 
\begin{figure}
\begin{center}
	\includegraphics[width=\linewidth, angle=0]{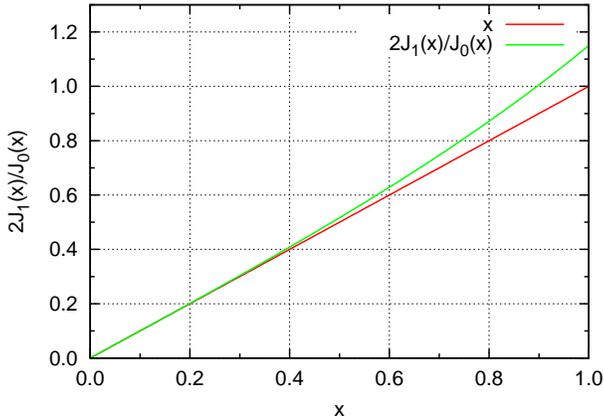} 
\end{center}
\caption{The ratio of $2J_1(x)/J_0(x)$ as a function of $x$. For $x \ll 1$, it is well approximated by $x$ as shown. 
For the whole range of $0\geq x \geq 1$, it is well approximated by a polynomial of the fifth order, which is derived from equation (\ref{eq:18}). From the value of $(A_{+m}+A_{-m})/A_0$, the value of $\alpha\xi_m$ satisfying equation (\ref{eq:43}) is obtained.}
\label{{fig:04}}
\end{figure}

By substituting the values of $\alpha\xi_m$ and $\theta_m$ into equation (\ref{eq:25}), we can formally reconstruct the radial velocity as a continuous function of time
from photometry alone:
\begin{eqnarray}
\lefteqn{
	v_{{\rm rad},\mathfrak{M}}(t) = -{{\Omega c}\over{\omega_0}} \sum_{m=1}^{\mathfrak{M}} m\alpha\xi_m\cos(m\Omega t+\theta_m)
}
\nonumber\\
\lefteqn{
\quad
	=
	-{{\Omega c}\over{\omega_0}} \sum_{m=1}^{\mathfrak{M}} m\alpha\xi_m\cos \left[ m\Omega \left(t+{{\theta_1}\over{\Omega}}\right)+(\theta_m-m\theta_1) \right],
}
\label{eq:46}
\end{eqnarray}
where $\mathfrak{M}$ is the number of components in the multiplet.   
In the limiting case of $\mathfrak{M} \rightarrow \infty$, this is the 
time-varying
radial velocity 
component, added to the systemic velocity, 
that would be measured by spectroscopy.
The number of terms, $\mathfrak{M}$, is, however, very limited and not large enough in practical cases, hence we need to devise a  more practical way of solving the problem. Nevertheless, equation (\ref{eq:46}) is instructive in the sense that it {\it proves that the radial velocity can be reproduced from the frequency spectrum of the photometric variations}.

\section{FM method}
\label{sec:4}
The amplitudes and phases of the multiplet components are determined by the binary orbit. 
Hence, inversely, once we have observed the frequency spectrum of a pulsating star in a binary system, we can extract information of the binary orbit. Essentially, this is the the FM method \citep{FM2012}.

While the strict values of the orbital elements can be obtained by analysing the continuous radial velocity, [equation (\ref{eq:46})], 
if the number of components of the multiplet $\mathfrak{M}$ is large enough, in practice $\mathfrak{M}$ is limited to $3$ or $4$. With such a limited number of terms, equation (\ref{eq:46}) may still substantially deviate from the real radial velocity. In order to improve this, we derive the orbital parameters first and reconstruct the radial velocity in the case of $\mathfrak{M}\rightarrow\infty$. It is then instructive to derive some `easy solutions' of the orbital parameters to illustrate the technique. More approximate solutions can be obtained directly from the Fourier transform of the light curve without carrying out further numerical computations. 
In this section, we treat such cases.

\subsection{The eccentricity, $\mbox{\boldmath$e$}$}
\label{sec:4.1}
\begin{figure}
\begin{center}
\includegraphics[width=\linewidth, angle=0]{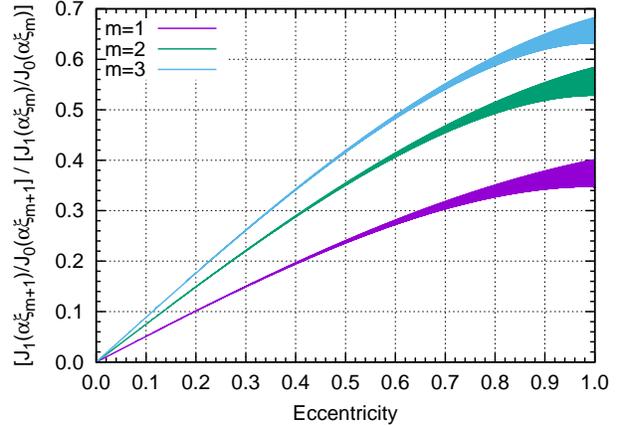} 
\caption{
${\cal R}_m(\alpha, e, \varpi)$, 
for a fixed value of $\alpha=0.217$, as a function of $e$. The width of each curve shows the range of $\varpi$ from $0$ to $2\upi$.
}
\label{fig:05}
\end{center}
\end{figure}

Equation (\ref{eq:43}) leads to the following relation between the amplitudes of the $m+1$-th and $m$-th sidelobes:
\begin{eqnarray}
\lefteqn{
	{{A_{+(m+1)}+A_{-(m+1)}}\over{A_{-m}+A_{-m}}} 
	= 
	{{J_1(\alpha\xi_{m+1})}\over{J_0(\alpha\xi_{m+1})}} \Big/ {{J_1(\alpha\xi_{m})}\over{J_0(\alpha\xi_{m})}} 
}
\label{eq:04new1}
\\
\lefteqn{
	\qquad\qquad\qquad\qquad\quad =: {\cal R}_m(\alpha, e, \varpi) .
}
\label{eq:04new2}
\end{eqnarray}
The RHS of equation (\ref{eq:04new1}), defined as ${\cal R}(\alpha, e, \varpi)$, 
is a function of $\alpha$, $e$ and $\varpi$, while the LHS is observable.
Fig.\,\ref{fig:05} shows ${\cal R}(\alpha, e, \varpi)$ as a function of $e$, for $m=1$, 2, and 3, for a fixed value of $\alpha=0.217$.
The width of each curve shows the range of $\varpi$ from $0$ to $2\upi$.

In the case of $\alpha\xi_m \ll 1$, $J_1(\alpha\xi_m) \simeq \alpha\xi_m/2$ and $J_0(\alpha\xi_m) \simeq 1$, hence
\begin{equation}
	{\cal R}_m \simeq {{\xi_{m+1}}\over{\xi_m}}  
\end{equation}
and then ${\cal R}$ is regarded as being no longer dependent on $\alpha$. It should be noted that, while $\{\xi_m\}$ are dependent on $e$ and $\varpi$, the ratios $\{\xi_{m+1}/\xi_m\}$, and then ${\cal R}_m$, are dependent mainly on $e$, but relatively weakly on $\varpi$ (see Fig.\,\ref{fig:05}). 
A good initial guess for the eccentricity is then derived from the amplitude ratio of the $(m+1)$-th sidelobes and $m$-th sidelobes.
Indeed, as seen in equation (\ref{eq:14}), the ratio $\xi_{m+1}/\xi_m$ is, in a good approximation, given by 
\begin{eqnarray}
\lefteqn{
	{{\xi_{m+1}}\over{\xi_m}} \simeq {{m}\over{m+1}}{{J_{m+1}((m+1)e)}\over{J_m(me)}} .
}
\label{eq:50}
\end{eqnarray}
Whether or not $\alpha\xi_m \ll 1$ can be checked by seeing the amplitude ratio of the sidelobe to the central component:
\begin{eqnarray}
\lefteqn{
	{{A_{+m}+A_{-m}}\over{A_0}} = 2{{J_1(\alpha\xi_m)}\over{J_0(\alpha\xi_m)}} 
}
\\
\lefteqn{
\qquad\qquad\qquad
	=: {\cal S}_m(\alpha\xi_m).
}
\label{eq:04new3}
\end{eqnarray}
If  $\alpha\xi_m \ll 1$, ${\cal S}_m(\alpha\xi_m) \simeq \alpha\xi_m \ll 1$.
So, if the amplitude ratio is substantially smaller than unity, the assumption of $\alpha\xi_m \ll 1$ is justified.
In such a case, the eccentricity is estimated as \citep{FM2012}
\begin{equation}
	e \simeq 2{{A_{+2}+A_{-2}}\over{A_{+1}+A_{-1}}}.
\label{eq:04new4}
\end{equation}

Use of equation (\ref{eq:50}) for more than two values of $m$ determines multiple values of $e$, each with its own uncertainty. The final value of $e$ should be determined as a weighted average of those, which, given that $A_m$ decreases as $m$ increases, will lead to the eccentricity being dominated by the $\xi_1$ and $\xi_2$ terms.

\subsection{The angle between the periapsis and the nodal point, $\mbox{\boldmath$\varpi$}$}
\label{sec:4.4}
The angles $\vartheta_1$ and $\vartheta_2$ are regarded as functions of $\varpi$ alone, since the eccentricity $e$ has already been obtained. 
The angle $\varpi$ between the periapsis and the nodal point is then derived from equation (\ref{eq:42}).

In the case of $e\ll 1$, as seen in Fig.\,2, the $e$-dependence of both $\vartheta_1$ and $\vartheta_2$ is weak, and 
$2\vartheta_1-\vartheta_2\simeq\varpi$.
Hence, in such a case,
\begin{equation}
	\varpi \simeq
	- {{1}\over{2}} \left[\phi_{+2} (t_{0}) -\phi_{-2} (t_{0})\right] +{{\upi}\over{2}}.
\label{eq:55}
\end{equation}
The uncertainty in $\phi_{+2} (t_{0}) -\phi_{-2} (t_{0})$ by $2n\upi$ can be eliminated by taking the inequality of $A_{\pm 1}$ 
and the deviation of $\phi_{\pm 1}(t_{0})-\phi_0$ from $\upi/2$ into account.

It should be noted that the angle $\varpi$ is obtained from information of the second sidelobes. 
If the amplitudes of $A_{\pm2}$ are not statistically significant,
we have to conclude that the orbit is close to a circular one.

\subsection{The projected semi-major axis, $\mbox{\boldmath$a_1\sin i$}$}
\label{sec:4.2}
By substituting the values of $e$ and $\varpi$ thus obtained into the first line of equation (\ref{eq:14}), we estimate $\xi_1(e, \varpi)$.
The value of $\alpha$ is then obtained by dividing $\alpha\xi_m$ (equation (\ref{eq:04new3})) by $\xi_m$. 
Then, the projected semi-major axis is derived by equation (\ref{eq:21}):
\begin{equation}
	a_1\sin i = {{\alpha}\over{\omega_0}} c.
\label{eq:53}
\end{equation}

\subsection{The mass function, $\mbox{\boldmath$f(m_1,m_2,\sin i)$}$}
\label{sec:4.3}
Once the value of 
$a_1\sin i$ 
is derived, with the help of the pulsation frequency and the orbital period, the mass function can be derived:
\begin{eqnarray}
\lefteqn{
	f(m_1,m_2,\sin i) := {{m_2^3\sin^3 i}\over{(m_1+m_2)^2}}
}
	\nonumber\\
\lefteqn{
\qquad\qquad\qquad\quad
	=
	\alpha^3 {{P_{\rm osc}^3}\over{P_{\rm orb}^2}} {{c^3}\over{2\upi G}} ,
}
\label{eq:54}
\end{eqnarray}
where $G$ is the gravitational constant. With a suitable asteroseismic estimate (or a reasonable assumption) of the primary mass, the minimum mass of the secondary star, $m_2 \sin i$, can be deduced.

\subsection{The radial velocity curve deduced from photometry}
\label{sec:4.5}
Once $e$ and $\varpi$ are determined, we can deduce $\cos f$ and $\sin f$ from equations (\ref{eq:11}) and (\ref{eq:12}), respectively, as functions of the mean anomaly, $\Omega(t-t_{\rm p})$, all the terms of $\xi_n(e,\varpi)$ and $\vartheta(e,\varpi)$.
Then substituting them, along with $a\sin i$ and $\Omega$, into equation (\ref{eq:7}), we obtain the radial velocity as a function of the orbital phase.

\subsection{Iteration for an improvement of parameters}
\label{sec:4.6}
Once the value of $\varpi$ is estimated, we may iterate the above mentioned processes, by taking account of the $\varpi$-dependences of $\xi_m$. The value of $\varpi$ newly derived in this way may be slightly different from that obtained previously. The iteration should be repeated until consistent solutions are obtained. An initial guess for $\xi$ is obtained by ignoring $\varpi$-dependence: $\xi_m^{(0)}=\xi(e^{(0)}, \varpi=0)$. The first iteration gives $\xi^{(1)}=\xi(e^{(0)}, \varpi^{(0)})$. We improve the value of $e$ as $e^{(1)}$ by adopting $\xi^{(1)}$, and then $\varpi^{(1)}$. We repeat these processes until we get consistent solutions.
Fig.\,\ref{fig:03} is not used in this process, but it is illustrative.

\begin{table*}
\centering
\caption[]{A least-squares fit of the frequency septuplet for the highest amplitude mode to the Q2--Q16 {\it Kepler} data for KIC\,9651065. The frequencies of the multiplet are separated by the orbital frequency, $\nu_{\rm orb} = 0.0036524 \pm 0.0000036$\,d$^{-1}$ ($P_{\rm orb} = 273.8 \pm 0.3$\,d). Column 4 shows that the first sidelobe phases are close to $\upi/2 = 1.57$\,rad out of phase with the central peak, as required by theory for FM. The zero point for the phases has been chosen to be a time when the phases of the first sidelobes are equal,  $t_0 = {\rm BJD}\,2455783.05262$, for the first multiplet. Column 5 shows that the phases of the first sidelobes of the other multiplets are equal within the errors at this time. }
\small
\begin{tabular}{ccrrrrr}
\toprule
\multicolumn{1}{c}{frequency} & 
\multicolumn{1}{c}{amplitude} &   
\multicolumn{1}{c}{phase} & 
\multicolumn{1}{c}{${\langle\phi_{\pm 1}\rangle}-{\phi_0}$} & 
\multicolumn{1}{c}{$\phi_{+m}-\phi_{-m}$} & 
\multicolumn{1}{c}{$\frac{A_{+m}+A_{-m}}{A_0}$} & 
\multicolumn{1}{c}{$\frac{A_{+m}-A_{-m}}{A_{+m}+A_{-m}}$} \\
\multicolumn{1}{c}{d$^{-1}$} & 
\multicolumn{1}{c}{mmag} &   
\multicolumn{1}{c}{rad} & 
\multicolumn{1}{c}{rad} & 
\multicolumn{1}{c}{rad} & & \\ 
\midrule
$19.46672$ & $0.0223 \pm 0.0020$ & $-1.4619 \pm 0.0896$ &   &   \\ 
$19.47037$ & $0.0613 \pm 0.0020$ & $0.7396 \pm 0.0326$ &   &   \\ 
$19.47402$ & $0.2213 \pm 0.0020$ & $-3.1416 \pm 0.0090$ &   &   \\ 
$19.47768$ & $1.9308 \pm 0.0020$ & $-1.5919 \pm 0.0010$ & $-1.550 \pm 0.007$ & &  &   \\ 
$19.48133$ & $0.2153 \pm 0.0020$ & $-3.1415 \pm 0.0093$ &  & $0.000 \pm 0.013$ & $0.2261 \pm 0.0029$ & $-0.014 \pm 0.006$\\
$19.48498$ & $0.0443 \pm 0.0020$ & $-0.4639 \pm 0.0452$ &  & $-1.204 \pm 0.056$ & $0.0547 \pm 0.0015$ & $-0.161 \pm 0.027$  \\ 
$19.48863$ & $0.0152 \pm 0.0020$ & $2.1830 \pm 0.1314$ &  & $3.645 \pm 0.132$ & $0.0194 \pm 0.0015$ & $-0.189 \pm 0.077$  \\ 
\midrule
$21.70118$ & $0.0096 \pm 0.0020$ & $0.5094 \pm 0.2090$ &   &   \\ 
$21.70483$ & $0.0332 \pm 0.0020$ & $2.5352 \pm 0.0602$ &   &   \\ 
$21.70848$ & $0.1163 \pm 0.0020$ & $-1.4206 \pm 0.0172$ &   &   \\ 
$21.71214$ & $0.8504 \pm 0.0020$ & $0.1291 \pm 0.0024$ & $-1.558 \pm 0.012$  &  & & \\ 
$21.71579$ & $0.1182 \pm 0.0020$ & $-1.4362 \pm 0.0169$ &  & $-0.016 \pm 0.024$ & $0.2758 \pm 0.0034$ & $0.008 \pm 0.012$\\
$21.71944$ & $0.0195 \pm 0.0020$ & $1.3598 \pm 0.1024$ &  & $-1.175 \pm 0.119$  & $0.0466 \pm 0.0033$ & $-0.346 \pm 0.076$ \\ 
$21.72309$ & $0.0064 \pm 0.0020$ & $-2.1510 \pm 0.3116$ &  & $-2.660 \pm 0.312$  & $0.0188 \pm 0.0033$ & $-0.200 \pm 0.180$ \\ 
\midrule
$30.79094$ & $0.0184 \pm 0.0020$ & $-1.8457 \pm 0.1088$ &   &   \\ 
$30.79459$ & $0.0398 \pm 0.0020$ & $-0.1224 \pm 0.0503$ &   &   \\ 
$30.79824$ & $0.1291 \pm 0.0020$ & $2.4607 \pm 0.0155$ &   &   \\ 
$30.80189$ & $0.7021 \pm 0.0020$ & $-2.3441 \pm 0.0028$ & $-1.516 \pm 0.011$ & &  &  \\ 
$30.80555$ & $0.1365 \pm 0.0020$ & $2.3847 \pm 0.0147$ &  & $-0.076 \pm 0.021$ & $0.3783 \pm 0.0042$ & $0.028 \pm 0.011$\\
$30.80920$ & $0.0301 \pm 0.0020$ & $-0.8970 \pm 0.0664$ &  & $-0.775 \pm 0.083$  &  $0.0996 \pm 0.0040$ & $-0.139 \pm 0.041$\\ 
$30.81285$ & $0.0075 \pm 0.0020$ & $0.4618 \pm 0.2682$ &  & $2.308 \pm 0.269$  &  $0.0369 \pm 0.0040$ & $-0.421 \pm 0.118$\\ 
\midrule
$36.13547$ & $0.0189 \pm 0.0020$ & $-2.1439 \pm 0.1061$ &   &   \\ 
$36.13913$ & $0.0389 \pm 0.0020$ & $-0.4995 \pm 0.0515$ &   &   \\ 
$36.14278$ & $0.0683 \pm 0.0020$ & $2.8955 \pm 0.0293$ &   &   \\ 
$36.14643$ & $0.2864 \pm 0.0020$ & $-1.9325 \pm 0.0070$ & $-1.484 \pm 0.022$ & &  &   \\ 
$36.15008$ & $0.0659 \pm 0.0020$ & $2.8373 \pm 0.0304$ &  & $-0.058 \pm 0.042$ & $0.4686 \pm 0.0104$ & $-0.018 \pm 0.021$\\
$36.15374$ & $0.0107 \pm 0.0020$ & $-0.7266 \pm 0.1873$ &  & $-0.227 \pm 0.194$ & $0.1732 \pm 0.0099$ & $-0.569 \pm 0.066$ \\ 
$36.15739$ & $0.0084 \pm 0.0020$ & $1.8010 \pm 0.2372$ &  & $3.945 \pm 0.239$ & $0.0953 \pm 0.0099$ & $-0.385 \pm 0.111$  \\ 
\bottomrule
\end{tabular}
\label{table:01}
\end{table*}

\section{Examples with real $\mbox{\boldmath${\it Kepler}$}$ data}
\label{sec:5}
The {\it Kepler} mission accumulated time series of photometric data of about 200\,000 stars from $2009 - 2013$. The long time base and high duty cycle of {\it Kepler} data make them the data of choice for FM analyses. Here we present examples for three highly eccentric \textit{Kepler} targets: KIC\,9651065, KIC\,10990452 and KIC\,8264492. In each case, we use multi-scale, maximum a posteriori (msMAP) pipeline data collected with long cadence (LC, 29.45-min) sampling.

\subsection{KIC\,9651065}
\label{sec:5.1}

\begin{figure}
\centering
\includegraphics[width=0.9\linewidth,angle=0]{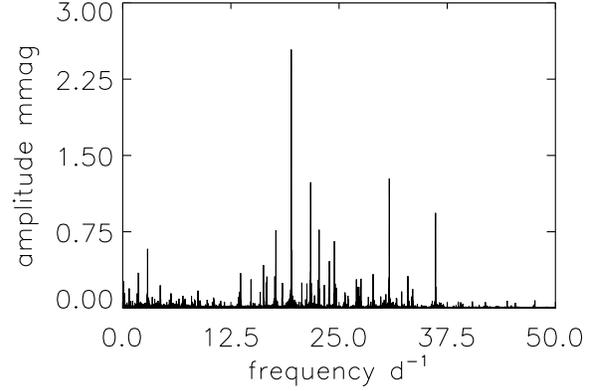} 
\caption{An amplitude spectrum for the Q5 SC data of KIC\,9651065. The four highest amplitude peaks were studied in LC. There are no significant peaks above 50\,d$^{-1}$. }
\label{{fig:06}}
\end{figure}

KIC\,9651065 is a $K_p = 11.1$ binary system showing first, second and third FM sidelobes, hence it has high eccentricity. 
According to \citet{Huber2014}, its effective temperature and gravity are $T_{\rm  eff} = 7014^{+128}_{-164}$\,K and $\log g = 3.833^{+0.132}_{-0.127}$ (cgs units), and it mass is $m_1=1.703^{+0.161}_{-0.185}$\,M$_{\odot}$, which we round to $T_{\rm eff} = 7400 \pm 150$\,K, $\log g = 3.83 \pm 0.13$ and $m_1=1.70 \pm 0.17$\,M$_{\odot}$  

\subsubsection{Kepler mission data}
\label{sec:5.1.1}
\begin{figure*}
\centering
\includegraphics[width=0.4\linewidth,angle=0]{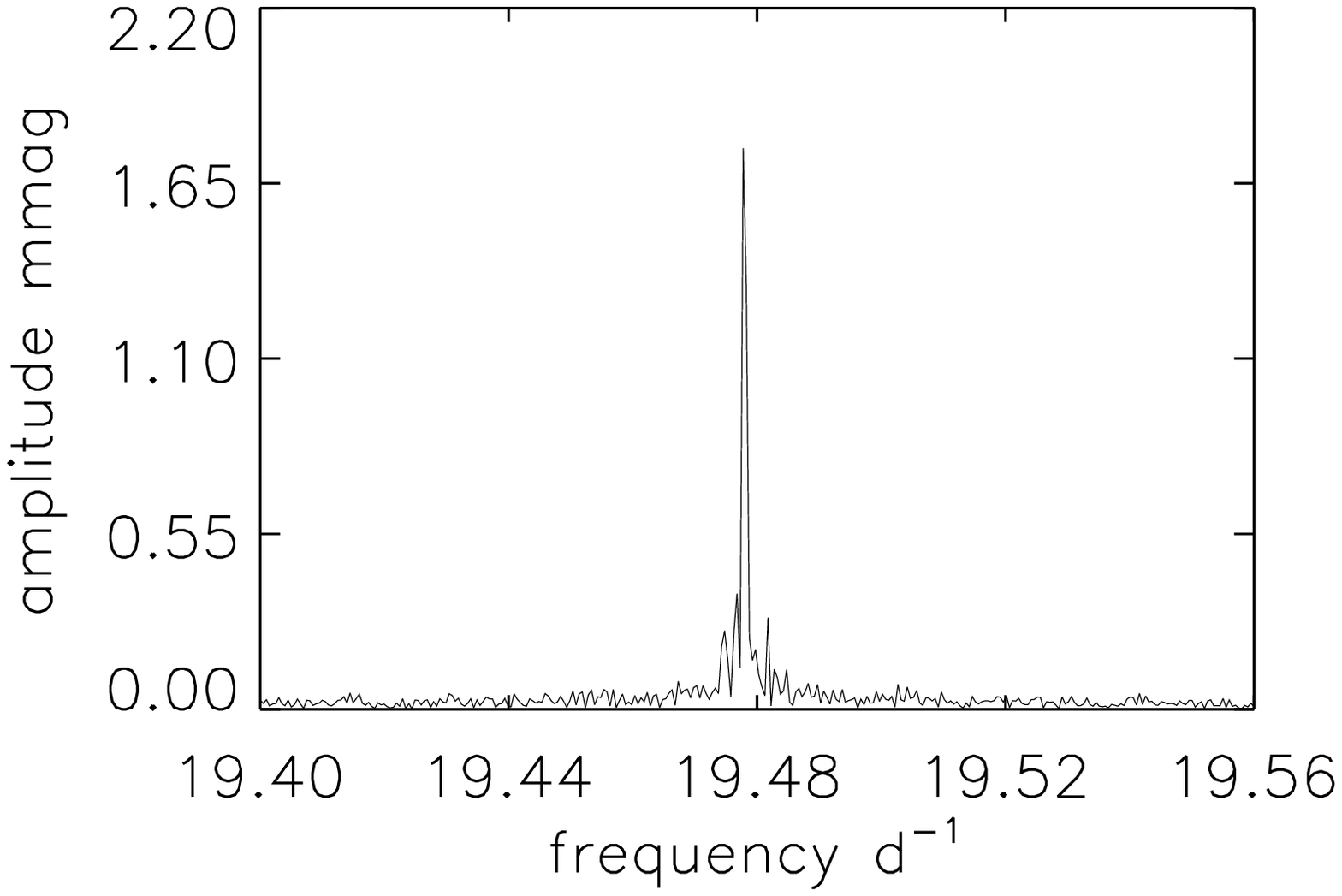} 
\includegraphics[width=0.4\linewidth,angle=0]{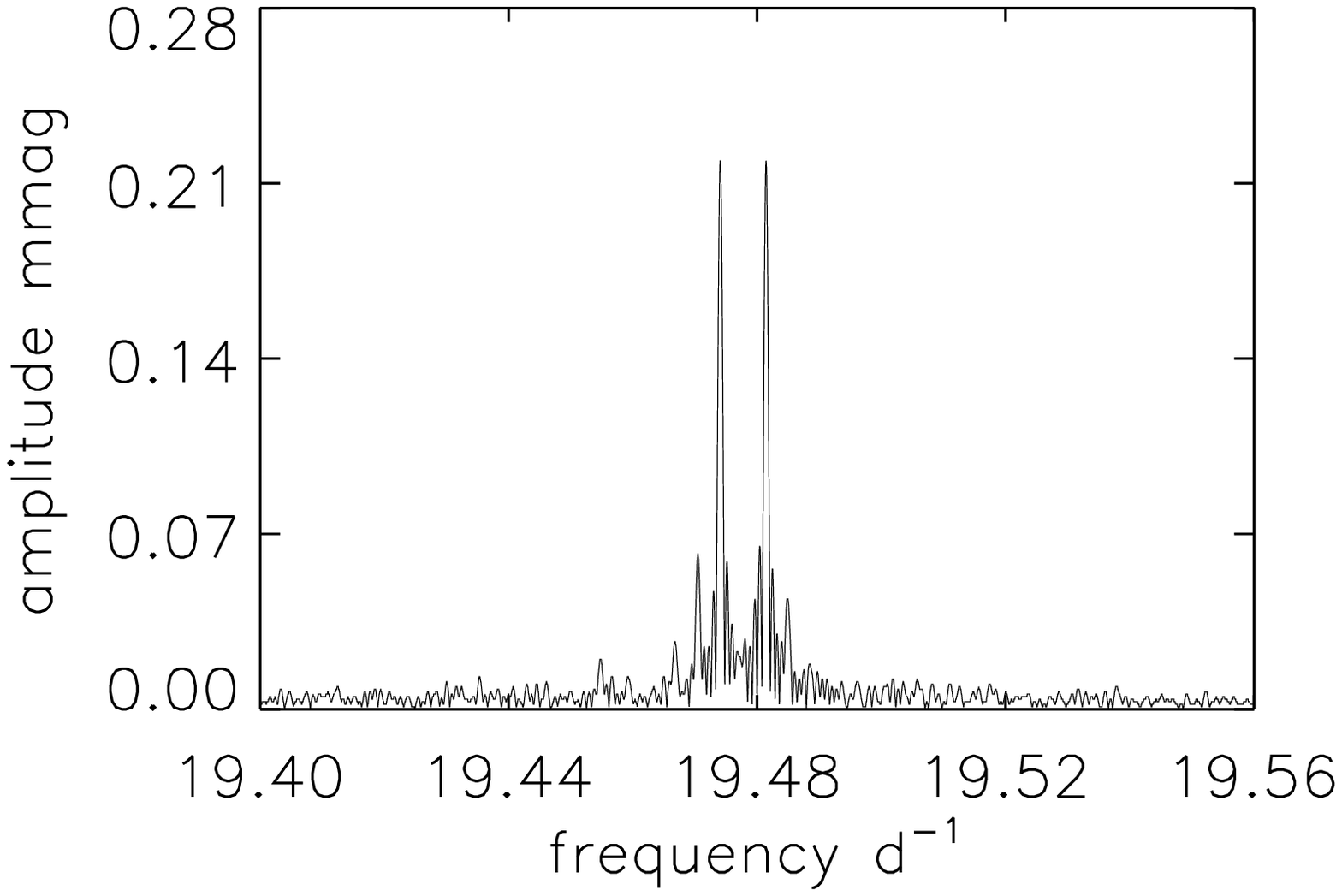} 
\includegraphics[width=0.4\linewidth,angle=0]{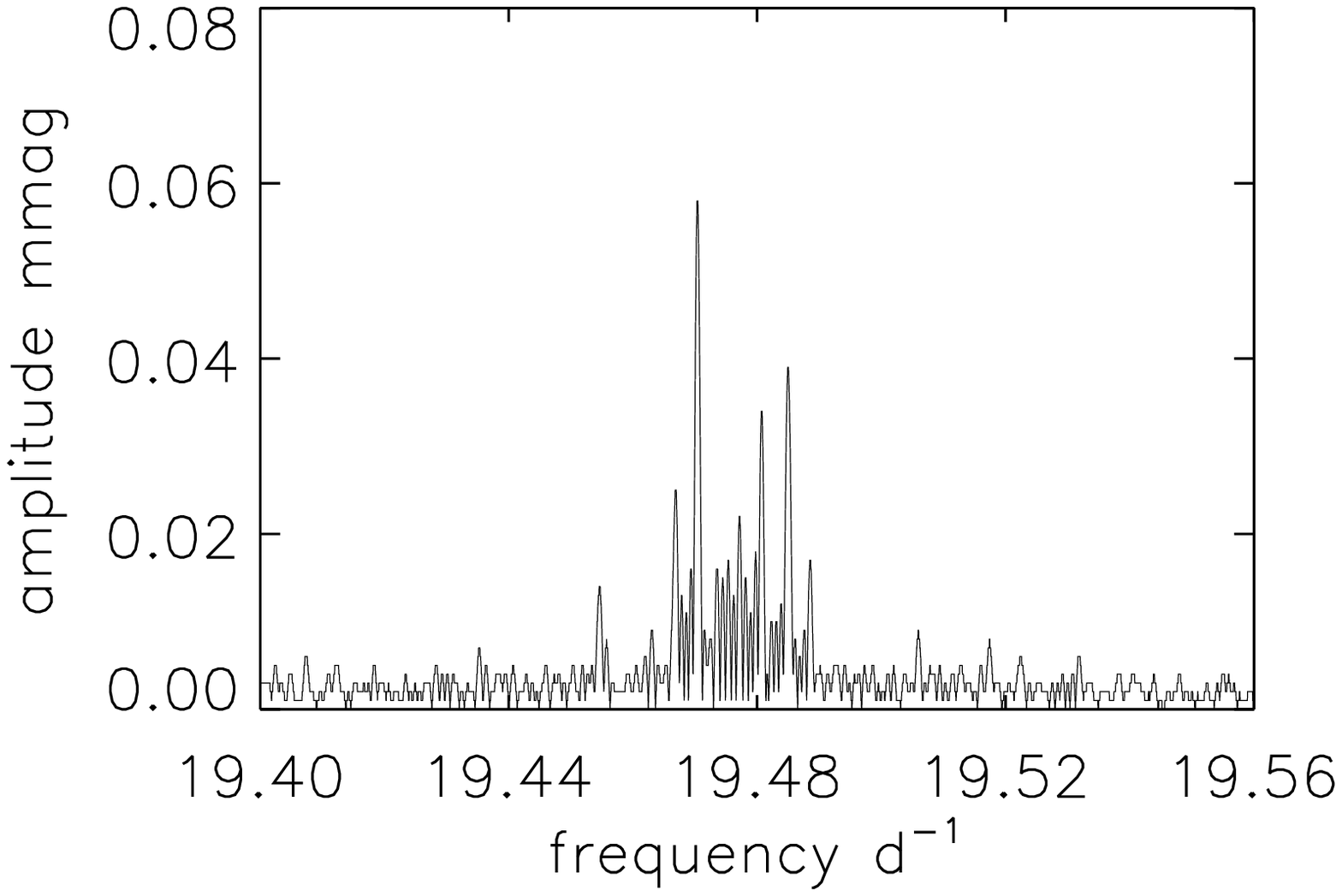} 
\includegraphics[width=0.4\linewidth,angle=0]{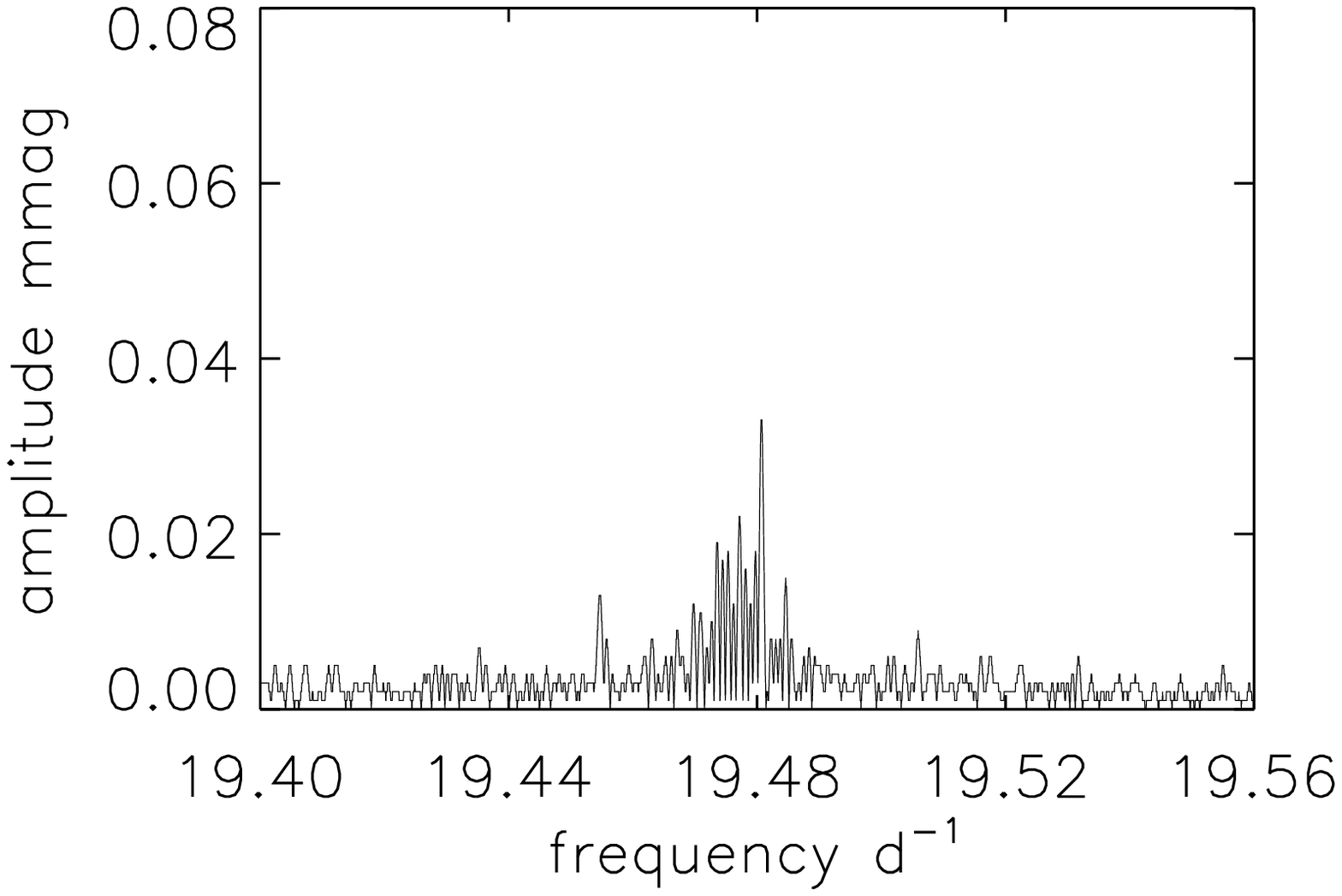} 
\caption{Amplitude spectra for KIC\,9651065 for the Q2--16 LC data. The top, left panel shows the highest amplitude pulsation peak at $\nu_1 = 19.47768$\,d$^{-1}$. The top, right panel shows the amplitude spectrum after prewhitening $\nu_1$ where two equally spaced sidelobes can be seen easily. These are the first FM sidelobes. A careful examination shows the second and third FM sidelobes also. The bottom left panel shows the amplitude spectrum after prewhitening $\nu_1$ and its first FM sidelobes, where the second and third sidelobes can be seen. The remaining variance close to the mode peak after pre-whitening the second and third sidelobes is seen in the bottom right panel; this is caused by amplitude modulation of the mode frequency over the time span of the data set, a common characteristic of $\delta$~Sct stars. The highest remaining peak may be that of an independent pulsation mode.} 
\label{{fig:07}}
\end{figure*}

The {\it Kepler} mission data available for this star are Q0--17 LC data, and Q5.1--5.3, Q7.1--7.3 short cadence (SC) data with 1-min integration time. An examination of the Q5 SC data shows that there are no significant peaks with amplitudes greater than 10\,$\umu$mag at frequencies higher than 60\,d$^{-1}$. In the frequency range $0-50$\,d$^{-1}$ the four highest amplitude peaks were studied for FM. Fig.\,\ref{{fig:06}} shows the amplitude spectrum for the Q5 SC data in the frequency range $0 - 50$\,d$^{-1}$ where the highest peaks chosen for analysis can be seen. They are in order of amplitude: $\nu_1 = 19.47768$\,d$^{-1}$, $\nu_2 = 30.80189$\,d$^{-1}$, $\nu_3 = 21.71214$\,d$^{-1}$ and $\nu_4 = 36.14643$\,d$^{-1}$. There is no clear separation of p\,mode and g\,mode frequency ranges; there are peaks in all frequency ranges, suggesting p\,modes, g\,modes and mixed modes. 

The Q0, Q1 and Q17 LC data were of shorter duration than full quarters, and they had a significant slope from the pipeline reductions. We therefore chose to analyse the Q2--16 LC data covering a time span of 1388.2\,d. We begin the discussion of the analysis with the highest amplitude peak at 19.47768\,d$^{-1}$. Fig.\,\ref{{fig:07}} shows the prewhitening process. The first, second and third FM sidelobes are significant and detected. There is some amplitude and/or frequency modulation of the $\nu_1$ mode, hence there is amplitude left near to that frequency after prewhitening. This is astrophysical, since amplitude and frequency variations on the time scale of this data set are common in $\delta$\,Sct stars (see, e.g., \citealt{bowman2014}).

This process was carried out for the other three of the highest four peaks. Figures are not shown for these other peaks. The results are given in Table\,\ref{table:01}. In the small sections of the amplitude spectrum between significant peaks the highest noise peaks have amplitudes of 8\,$\umu$mag. The error on the amplitude determination is taken to be one quarter of that, with phase and frequency errors scaled accordingly \citep{montgomery99}. The formal least-squares errors contain variance from all the peaks, so must be scaled down by this estimate.

\subsubsection{FM analysis}
\label{sec:5.1.2}
For all four modes presented in Table\,\ref{table:01} all sidelobes are equally split from the central frequency, being consistent with theoretical expectation of FM stars. This splitting gives the orbital frequency, $\nu_{\rm orb} = 0.0036524 \pm 0.0000036$\,d$^{-1}$ ($P_{\rm orb} = 273.8 \pm 0.3$\,d), where we have taken the best value here to be from the splitting of the first sidelobes of the highest amplitude mode. The zero point for the phases has been chosen to be a time when the phases of the first sidelobes of the highest amplitude peak are equal. The phases of the first sidelobes are close to $\upi/2$\,rad out of phase with the central peak. That they are slightly less than this is a consequence of the values of the eccentricity and argument of periapsis. The same is true for the other modes within the error bars. Furthermore, the phase relation $(\phi_{+n}-\phi_{-n})$ for $n=2$, $3$, $4$ is the same within the errors for the $\nu_1$ and $\nu_2$ modes. As for the $\nu_3$ and $\nu_4$ modes, the uncertainty is too large to confirm this. These facts are consistent with the theory for FM discussed in Section 3.3.  

\begin{table}
\centering
\caption[]{The estimated quantities of the terms for KIC\,9651065, $m\alpha\xi_m$ and $\theta_m$, appearing in the expression of the radial velocity. Column 1 shows the frequency of the central peak, and Column 2 shows its ratio to the orbital frequency. Here $c$ is the speed of light. 
Column 4 shows $m\alpha\xi_m$ estimated from equation (\ref{eq:43}). 
Column 5 shows $\theta_m-m\theta_1$ estimated by equation (\ref{eq:45}). 
}
\small
\begin{tabular}{ccccr}
\toprule
\multicolumn{1}{c}{$\nu_{\rm puls}$} & 
\multicolumn{1}{c}{$(\nu_{\rm orb}/\nu_{\rm puls})c$} & 
\multicolumn{1}{c}{$m$} & 
\multicolumn{1}{c}{$m\alpha\xi_m$} & 
\multicolumn{1}{c}{$\theta_m-m\theta_1$} \\ 
\multicolumn{1}{c}{d$^{-1}$} & 
\multicolumn{1}{c}{km\,s$^{-1}$} &   
 & 
 & 
\multicolumn{1}{c}{rad}\\ 
\midrule
$19.47768$ & $57.0 \pm 0.2$ & $1$ & $0.2247 \pm 0.0015$ & $ 0.000 \pm 0.007$ \\ 
                    &                          & $2$ & $0.1094 \pm 0.0030$ & $ 0.969 \pm 0.028$ \\ 
                    &                          & $3$ & $0.0582 \pm 0.0045$ & $ 4.964 \pm 0.066$ \\ 
\midrule
$21.71214$ & $51.1 \pm 0.2$ & $1$ & $0.2732 \pm 0.0033$ & $-0.008 \pm 0.012$ \\ 
                    &                          & $2$ & $0.0932 \pm 0.0066$ & $ 0.983 \pm 0.059$ \\ 
                    &                          & $3$ & $0.0564 \pm 0.0099$ & $ 4.953 \pm 0.156$ \\ 
\midrule
$30.80189$ & $36.1 \pm 0.1$ & $1$ & $0.3717 \pm 0.0040$ & $-0.038 \pm 0.011$ \\
                    &                          & $2$ & $0.1992 \pm 0.0080$ & $ 1.184 \pm 0.042$ \\
                    &                          & $3$ & $0.1107 \pm 0.0120$ & $ 4.295 \pm 0.134$ \\
\midrule
$36.14643$ & $30.7 \pm 0.1$ & $1$ & $0.4563 \pm 0.0096$ & $-0.029 \pm 0.021$ \\
                    &                          & $2$ & $0.3452 \pm 0.0194$ & $ 1.457 \pm 0.097$ \\
                    &                          & $3$ & $0.2856 \pm 0.0297$ & $ 5.114 \pm 0.120$ \\
\bottomrule
\end{tabular}
\label{table:02}
\end{table}

For each set of sidelobes, the values of $m\alpha\xi_m$ and $\theta_m-m\theta_1$ that were derived by equations (\ref{eq:43}) and (\ref{eq:45}), respectively, are summarized in Table\,\ref{table:02}. The eccentricity $e$ is then estimated from the first  and second sidelobes around $\nu_1$ by equation (\ref{eq:04new4}) to be $e = 0.569 \pm 0.030$.  The technique explained in Section 4.1 and inspection of Fig.\,\ref{fig:03} gives the coefficient $\xi_1(e, \varpi) = 0.840\pm 0.015$. Using this value and the amplitude ratio for $\nu_1$, $\alpha\xi_1=0.2247\pm 0.0015$, we obtain $\alpha = 0.2675\pm 0.0018$. The mass function is then estimated to be $0.0992 \pm 0.0057$, which gives a minimum secondary mass of $m_2 = 0.87\pm 0.02$\,M$_{\odot}$ based on an assumption of $m_1=1.7$\,M$_{\odot}$, so the secondary is probably a main sequence G star. We derive the semi-major axis of the primary star about the barycentre from equation (\ref{eq:21}) to be $a_1 \sin i = 0.378 \pm 0.007$\,au from the data for $\nu_1$. 

The facts that  $A_{+2}-A_{-2} <0$ and $\langle\phi_{\pm 1}\rangle-\phi_0 > -\upi/2$ indicate $0 <  2\vartheta_1 -\vartheta_2 < \upi$.
The phase difference between $\phi_{+2}$ and $\phi_{-2}$, along with this constraint, leads to $2\vartheta_1-\vartheta_2 =\mbox{$2.17\pm 0.03$}$\,rad.
With use of the derived value of $e$, this gives $\varpi=2.22 \pm 0.04$\,rad.

\begin{figure}
\begin{center}
	\includegraphics[width=0.9\linewidth, angle=0]{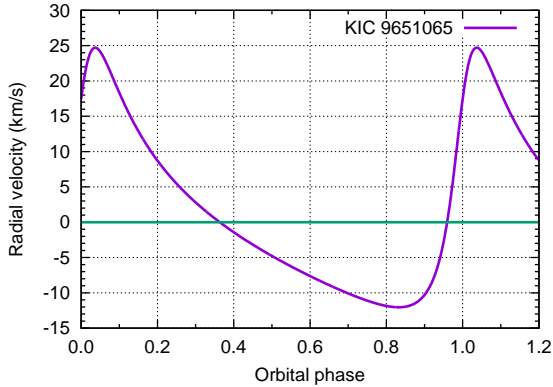} 
\end{center}
\caption{
The radial velocity curve, derived from the Kepler photometric data with the FM method, of KIC\,9651065. The zero point for the orbital phase corresponds to the periapsis passage of the star.} 
\label{fig:08}
\end{figure}

Fig.\,\ref{fig:08} shows the radial velocity curve of KIC\,9651065 derived from the {\it Kepler} photometric data alone. 
This demonstrates again that the present FM method extracts binary information from the {\it Kepler} light curve without spectroscopic observations.

\subsubsection{Comparison with the result obtained by PM}
\label{sec:5.1.3}
\cite{PM2014} developed a different method, by which the phase modulation (PM) in the time domain of intrinsic pulsation frequencies of the star is tracked. 
In this method, first of all, the frequency of the central component of the multiplet is measured in the Fourier transform. Then the light curve is divided into short segments and, with the frequency fixed, the pulsation phase in each segment is measured by a least-squares method. This provides us with ``time delays'' $\tau(t)$ as a function of time, where $\tau(t)$ is defined as in equation (\ref{eq:1}). Then by carrying out a Fourier analysis, we obtain the amplitudes $\alpha\omega_0^{-1}\xi_n(e,\varpi)$ and phases $\vartheta_n(e, \varpi)$ given in equation (\ref{eq:19}) for $n=1, 2, 3, \cdots$.
The amplitude ratios between $\xi_n$ and $\xi_{n+1}$ allow us to estimate the eccentricity $e$. On the other hand, the radial velocity is deduced as a function of time by taking the time derivative of the time delay,
\begin{equation}
	v_{\rm rad}(t) =  c
	{{{\rm d}\tau}\over{{\rm d}t}}.
\label{eq:56}
\end{equation}
Then, from the maximum and minimum radial velocity, along with the eccentricity deduced from the amplitude ratios, the projected semi-major axis is deduced;
\begin{equation}
	a_1\sin i = {{1}\over{2\Omega}} \sqrt{1-e^2}(v_{\rm rad, max}-v_{\rm rad, min}).
\label{eq:57}
\end{equation}
The angle $\varpi$ is also deduced from $v_{\rm rad, max}$ and $v_{\rm rad, min}$:
\begin{equation}
	\cos\varpi = -{{1}\over{e}} 
	\left[
	{{v_{\rm rad, max}+v_{\rm rad, min}}\over{v_{\rm rad,  max}-v_{\rm rad, min}}}
	\right] .
\label{eq:58}
\end{equation}

\cite{PM2014} applied this PM method to several pulsating stars observed by the Kepler Mission, including KIC\,9651065. They derived the orbital elements of this star, as well as the radial velocity curve. The results obtained in this paper by the FM method and their PM results are in good agreement (see Table\,\ref{table:03}). 

It should be noted here that the PM method has been further developed by \cite{PM2}. The latest method derives directly the orbital elements without converting time delays to radial velocities, offering higher precision.

\begin{table}
\centering
\caption[]{Comparison of the binary parameters of KIC\,9651065 derived with the FM method and the PM method. 
}
\small
\begin{tabular}{cr@{\,$\pm$\,}lr@{\,$\pm$\,}l}
\toprule
\multicolumn{1}{c}{Quantity}& 
\multicolumn{2}{c}{PM} & 
\multicolumn{2}{c}{FM} \\
\midrule
$P_{\rm orb}$ (d) & $272.70$ &  $0.82$  & $273.80$ & $0.30$ \\
$e$                       & $0.47$     & $0.03$   & $0.57$    & $0.03$ \\ 
$\varpi$  (rad)       & $2.01$     & $0.30$   & $2.22$   & $0.04$ \\
$a_1\sin i$  (au)        & $0.37$     & $0.02$   & $0.38$   & $0.01$ \\
$f(m_1,m_2,\sin i)$ (M$_\odot$) & $0.0916$ & $0.0108$ & $0.0992$ & $0.0057$ \\
\bottomrule
\end{tabular}
\label{table:03}
\end{table}

\subsection{KIC\,10990452}
\label{sec:5.2}

\begin{figure}
\centering
\includegraphics[width=0.9\linewidth,angle=0]{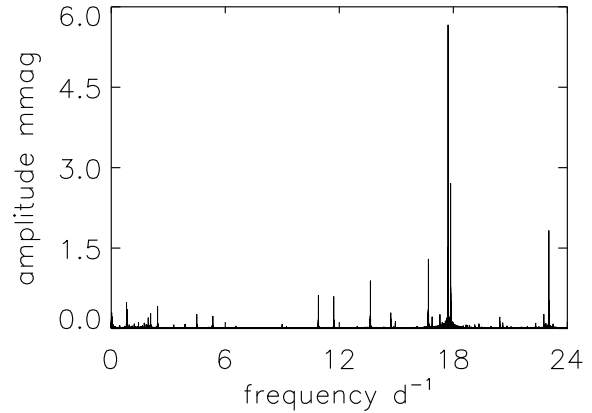} 
\caption{Amplitude spectrum of the LC Q1-16 data for KIC\,10990452. Most of the variance is in the p\,mode frequency range with a dominant highest peak at $\nu_1=17.72374$\,d$^{-1}$. } 
\label{fig:09}
\end{figure}

KIC\,10990452 is a $K_p = 12.4$ binary system showing first, second, third and one component of the fourth FM sidelobes, indicating that the system is highly eccentric. According to \citet{Huber2014}, its effective temperature and gravity are $T_{\rm eff} = 7585^{+244}_{-287}$\,K and $\log g = 4.022^{+0.126}_{-0.316}$ (cgs units), and its mass is $m = 1.691^{+0.332}_{-0.219}$\,M$_{\odot}$, which we round to $T_{\rm eff} = 7600 \pm 250$\,K, $\log g = 4.02 \pm 0.13$ and $m_1=1.69 \pm 0.33$\,M$_{\odot}$. These are consistent with the original KIC parameters.  We used Q1--16 data in our analysis, which have gaps because the star fell on the failed Module 3. Light curves in quarters Q7, Q11 and Q15 are therefore missing. The short 9.7-d engineering Q0 and the short 31-d final  Q17 were not included.

\begin{figure*}
\centering
\includegraphics[width=0.45\linewidth,angle=0]{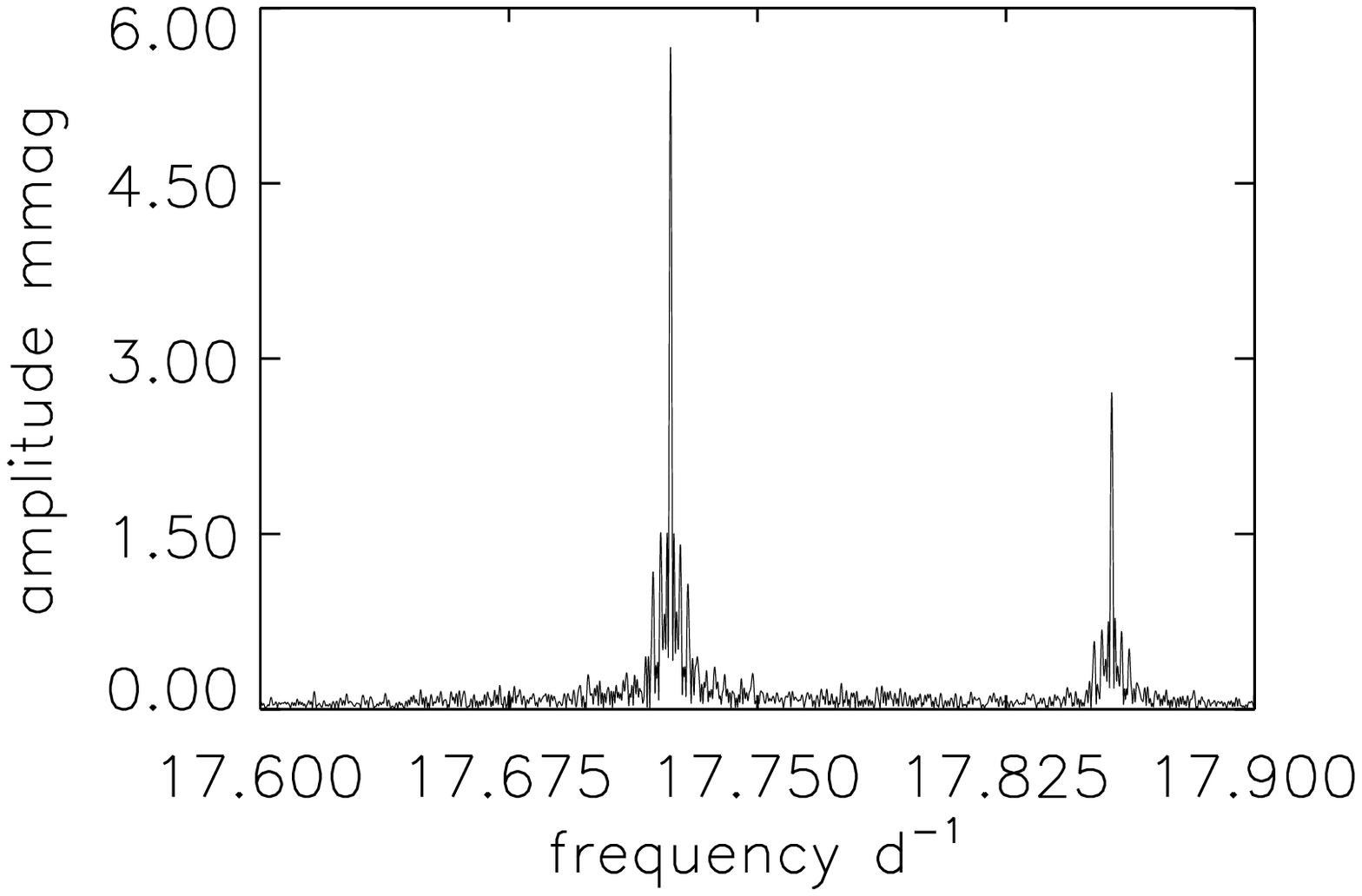} 
\includegraphics[width=0.45\linewidth,angle=0]{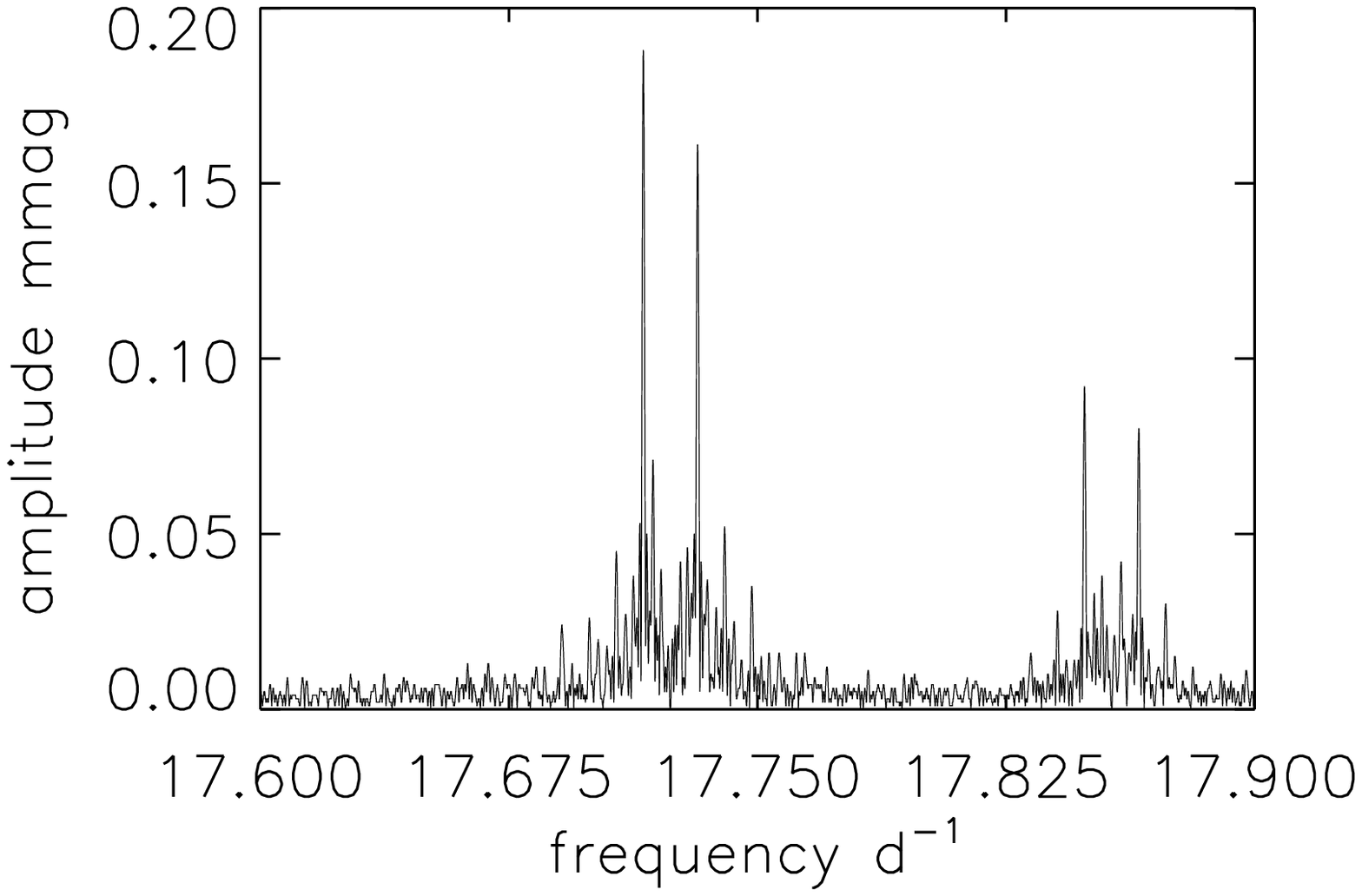} 
\includegraphics[width=0.45\linewidth,angle=0]{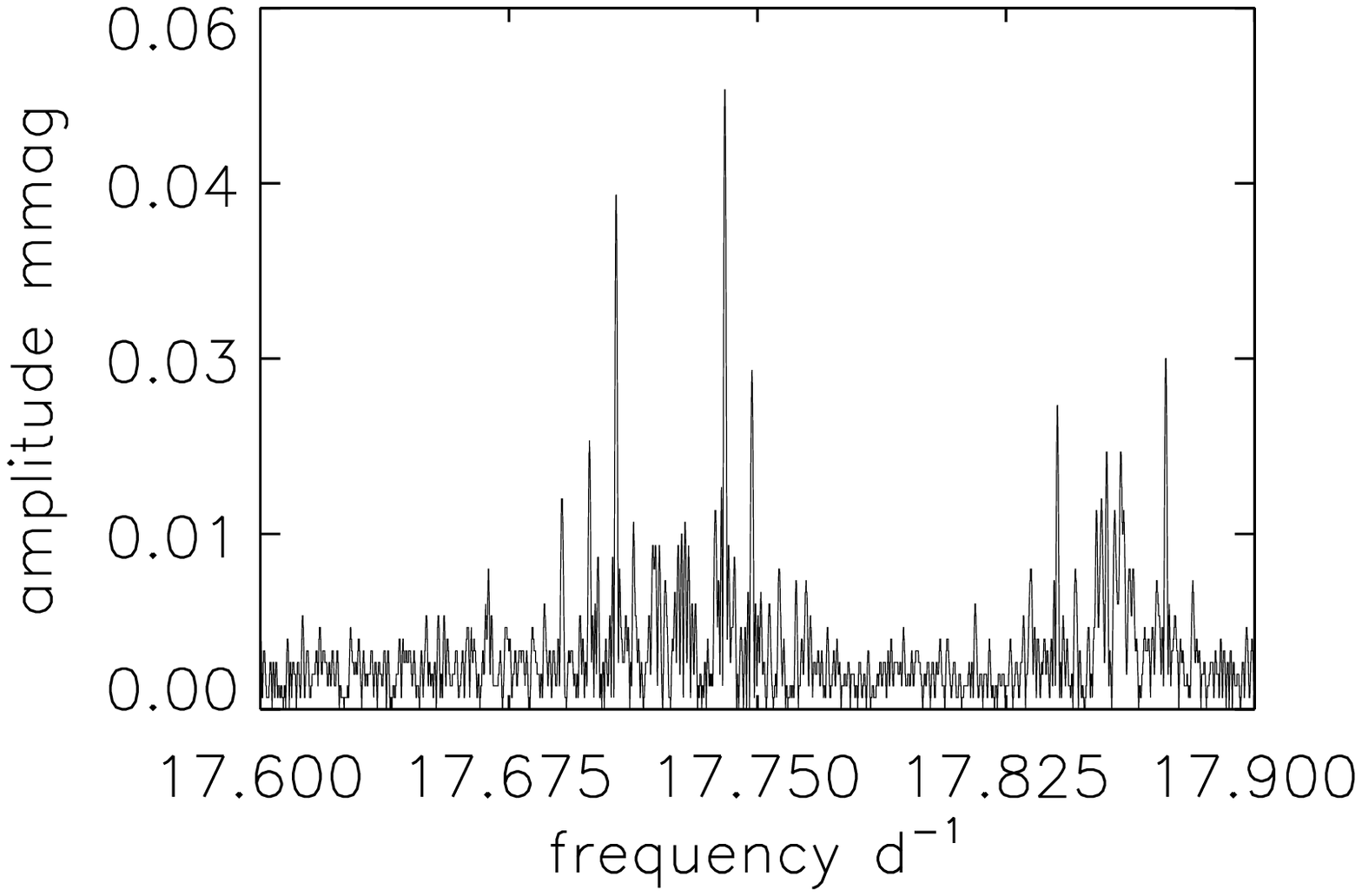} 
\includegraphics[width=0.45\linewidth,angle=0]{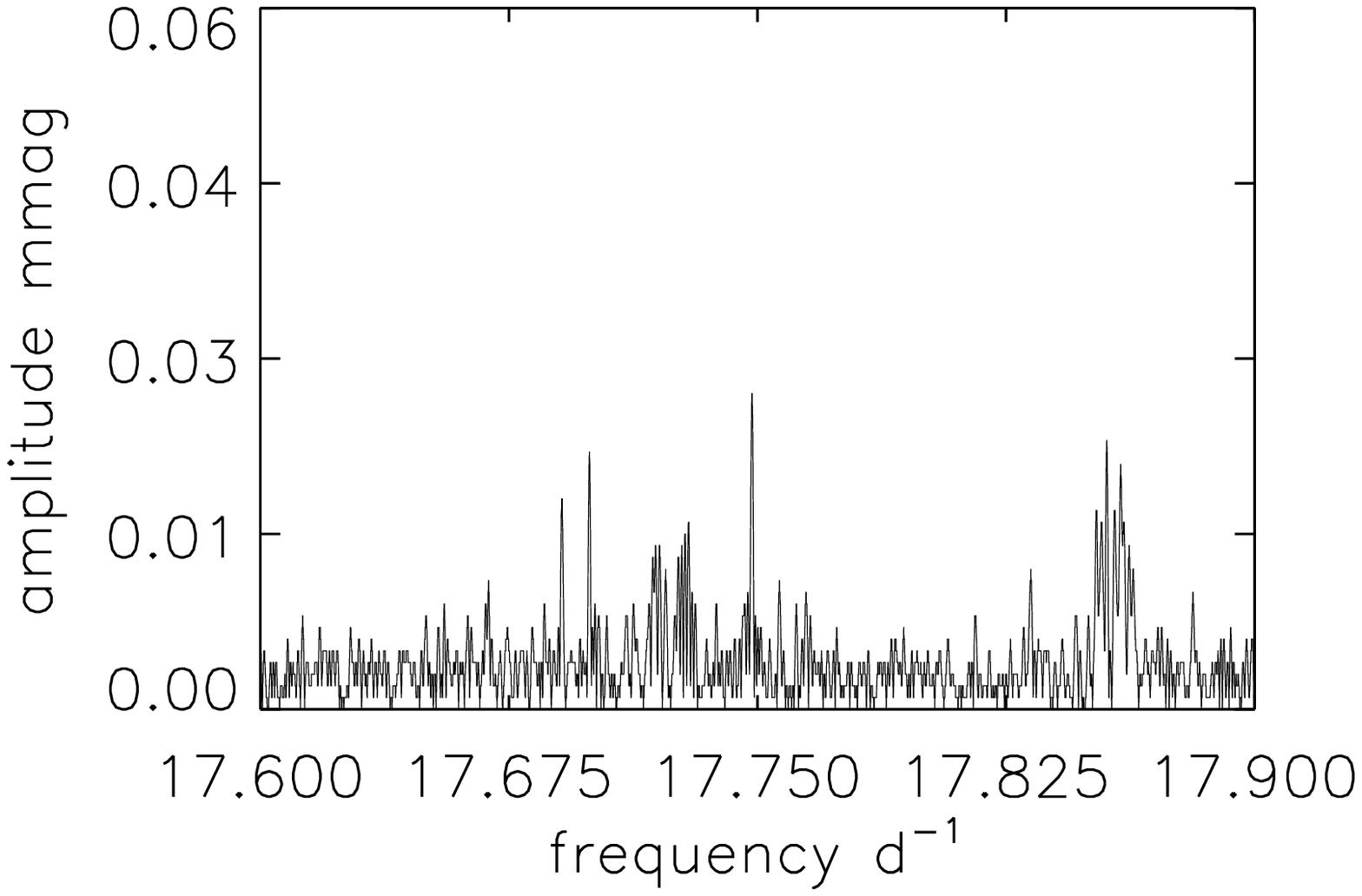} 
\caption{Top left: Amplitude spectrum for KIC\,10990452 in the frequency range of the two highest amplitude p-mode peaks. The spectral window has significant sidelobes because of the missing quarters of data. Top right: This panel shows the first FM sidelobes after prewhitening by $\nu_1$ and $\nu_2$; further sidelobes can be seen. Bottom left: The this panel shows the data after prewhitening by the first sidelobes, too, where the second and third sidelobes are easily visible. Bottom right: In this panel the third sidelobes are easily seen for both main frequencies, and the low frequency fourth sidelobe is also seen for the $\nu_1$ group. }
\label{fig:10}
\end{figure*}

Fig.\,\ref{fig:09} shows the LC amplitude spectrum where it can be seen that there is a dominant highest peak. We examine that and the second highest amplitude peak next to it for FM. The top left panel of Fig.\,\ref{fig:10} shows the highest peak at $\nu_1=17.72374$\,d$^{-1}$ and the second highest peak close to it, $\nu_2=17.85683$\,d$^{-1}$. Because of the complexity of the spectral window caused by the gaps in the data, the window patterns for the two frequencies interfere, so they must be modelled together. The top right panel shows the amplitude spectrum after the highest two mode peaks have been prewhitened; the FM sidelobes are apparent for both modes. The middle left panel is one step further with prewhitening of the first sidelobes, too. The remaining central amplitude of $\nu_2$ represents amplitude modulation over the time span of the data set -- either astrophysical or instrumental. Given that $\nu_1$ does not show this as much, it is probably astrophysical. The middle right panel shows the third sidelobes and the low-frequency component of the fourth sidelobes. The bottom panel shows the amplitude spectrum of the residuals after the two nonuplets given in Table\,\ref{table:04} have been prewhitened.

\begin{table*}
\centering
\caption[]{A least-squares fit of the frequency nonuplets for the two highest amplitude modes to the Q1 to Q16 {\it Kepler} data for KIC\,10990452. 
The frequencies of the multiplet are separated by the orbital frequency, $\nu_{\rm orb} = 0.008190 \pm 0.000024$\,d$^{-1}$ ($P_{\rm orb} = 122.11 \pm 0.36$\,d). 
The zero point for the phases has been chosen to be a time when the phases of the first sidelobes for the highest amplitude peak are equal,  $t_0 = {\rm BJD}\,2455673.01863$. 
Column 4 shows that the first sidelobe phases are $\upi/2 = 1.57$\,rad out of phase with the central peak, as required by theory for FM. 
Column 5 shows that the phases of the first sidelobes are equal within the errors at this time.
}
\small
\begin{tabular}{ccrrrcr}
\toprule
\multicolumn{1}{c}{frequency} & 
\multicolumn{1}{c}{amplitude} &   
\multicolumn{1}{c}{phase} & 
\multicolumn{1}{c}{${\langle\phi_{\pm 1}\rangle}-{\phi_0}$} & 
\multicolumn{1}{c}{$\phi_{+m}-\phi_{-m}$} & 
\multicolumn{1}{c}{$\frac{A_{+m}+A_{-m}}{A_0}$} & 
\multicolumn{1}{c}{$\frac{A_{+m}-A_{-m}}{A_{+m}+A_{-m}}$} \\ 
\multicolumn{1}{c}{d$^{-1}$} & 
\multicolumn{1}{c}{mmag} &   
\multicolumn{1}{c}{rad} & 
\multicolumn{1}{c}{rad} & 
\multicolumn{1}{c}{rad} & &  \\ 
\midrule
$17.6909790$ & $0.018 \pm 0.003$ & $ 0.7772 \pm 0.1691$ &   &   &   &   \\ 
$17.6991685$ & $0.021 \pm 0.003$ & $-0.1864 \pm 0.1450$ &   &   &   &   \\ 
$17.7073581$ & $0.044 \pm 0.003$ & $-1.0048 \pm 0.0684$ &   &   &   &   \\ 
$17.7155476$ & $0.194 \pm 0.003$ & $-2.1688 \pm 0.0154$ &   &   &   &   \\ 
$17.7237371$ & $5.654 \pm 0.003$ & $-0.6296 \pm 0.0005$ & $-1.539 \pm 0.012$  &  & &   \\ 
$17.7319266$ & $0.168 \pm 0.003$ & $-2.1688 \pm 0.0179$ &  & $0.000 \pm 0.024$ & $0.064 \pm 0.001$ & $-0.072 \pm 0.012$ \\
$17.7401161$ & $0.052 \pm 0.003$ & $ 3.1176 \pm 0.0577$ &  & $4.122 \pm 0.090$  & $0.017 \pm 0.001$ & $0.083 \pm 0.044$ \\ 
$17.7483056$ & $0.027 \pm 0.003$ & $ 1.6975 \pm 0.1114$ &  & $1.884 \pm 0.113$  & $0.008 \pm 0.001$ & $0.125 \pm 0.089$ \\ 
$17.7564952$ & $0.010 \pm 0.003$ & $ 0.4999 \pm 0.3118$ &  & $-0.277 \pm 0.312$  & $0.005 \pm 0.001$ & $-0.286 \pm 0.158$ \\  
\midrule
$17.8240752$ & $0.002 \pm 0.003$ & $-2.5761 \pm 1.3006$ &   &   &   &   \\ 
$17.8322647$ & $0.011 \pm 0.003$ & $2.1708 \pm 0.2699$ &   &   &   &   \\ 
$17.8404542$ & $0.026 \pm 0.003$ & $1.6322 \pm 0.1155$ &   &   &   &   \\ 
$17.8486437$ & $0.095 \pm 0.003$ & $0.4765 \pm 0.0317$ &   &   &   &   \\ 
$17.8568332$ & $2.674 \pm 0.003$ & $1.9836 \pm 0.0011$ &  $-1.501 \pm 0.024$ &  &  &   \\ 
$17.8650228$ & $0.083 \pm 0.003$ & $0.4883 \pm 0.0362$ &  & $0.012 \pm 0.048$ & $0.067 \pm 0.002$ & $-0.067 \pm 0.024$\\ 
$17.8732123$ & $0.029 \pm 0.003$ & $-0.6460 \pm 0.1022$ &  & $-2.278 \pm 0.154$ & $0.021 \pm 0.002$ & $0.055 \pm 0.077$ \\ 
$17.8814018$ & $0.009 \pm 0.003$ & $-1.8700 \pm 0.3233$ &  & $-4.041 \pm 0.325$ & $0.007 \pm 0.002$  & $-0.100 \pm 0.213$ \\ 
$17.8895913$ & $0.004 \pm 0.003$ & $-2.4669 \pm 0.6935$ &  & $0.109 \pm 0.694$  & $0.002 \pm 0.002$  & $0.333 \pm 0.745$ \\ 
\bottomrule
\end{tabular}
\label{table:04}
\end{table*}

When searching for significant peaks associated with pulsation modes among thousands of independent frequencies in the amplitude spectrum -- as we do in this case -- it is found that the highest noise peaks have amplitudes about 4 times that of the rms noise in amplitude. The least-squares fitting that we do to find frequencies, amplitudes and phases estimates the errors based on the total variance in the data. Since there are so many other pulsation frequencies present that have not been modelled, the noise is not white and errors are significantly overestimated. We therefore looked at a section of the amplitude spectrum that is pure noise and found that the highest noise peaks are around 12\,$\umu$mag. The amplitude error was therefore scaled to one quarter of that, or 3\,$\umu$mag. The phase and frequency errors were scaled by the same factor, since the errors on those quantities are proportional to the amplitude signal-to-noise ratio \citep{montgomery99}.

The results of the least-squares fits of the two pulsation mode peaks and their first, second, third, and fourth FM sidelobes are given in Table\,\ref{table:04}. All sidelobes are equally split from the central frequency, and is consistent with theoretical expectation for FM stars. This splitting gives the orbital frequency, $\nu_{\rm orb} = 0.008190 \pm 0.000024$\,d$^{-1}$ ($P_{\rm orb} = 122.11 \pm 0.36$\,d). The zero point for the phases has been chosen to be a time when the phases of the first sidelobes of the highest amplitude peak are equal. The phases of the first sidelobes are $\upi/2$\,rad out of phase with the central peak.
The same is true for the second highest amplitude mode with $\nu_2$. Furthermore, the phase relation $(\phi_{+m}-\phi_{-m})$ for $m=2$, $3$, $4$ is the same within the errors for the two modes. These facts are consistent with theory for FM discussed in Section 3.3.  

For each set of sidelobes, the values of $m\alpha\xi_m$ and $\theta_m-m\theta_1$ were derived by equations (\ref{eq:43}) and (\ref{eq:45}), respectively, and are summarized in Table\,\ref{table:05}. The values of these quantities derived from $\nu_1$ and $\nu_2$ are in good agreement, and this is consistent with theoretical expectation of FM stars. The eccentricity $e$ is then estimated from the first  and second sidelobes around $\nu_1$ by equation (\ref{eq:04new4}) to be $e = 0.569 \pm 0.030$. 

The technique explained in Section 4.1 and inspection of Fig.\,\ref{fig:03} gives the coefficient $\xi_1(e, \varpi) = 0.815\pm 0.015$. Using this value and the amplitude ratio for $\nu_1$, $\alpha\xi_1=0.0640\pm 0.0015$, we obtain $\alpha = 0.0785\pm 0.002$. The mass function is determined from these values to be $f(m_1,m_2,\sin i)=0.0163\pm 0.0011\,{\rm M}_{\odot}$. Assuming $m_1=1.7$\,M$_\odot$ \citep{Huber2014} gives a minimum secondary mass of $m_2 = 0.41\pm 0.01$\,M$_{\odot}$, so the secondary is probably a main sequence M or K star. We also derive the semi-major axis of the primary star about the barycentre. From equation (\ref{eq:21}), we find $a_1 \sin i = 0.122 \pm 0.003$\,au from the data for $\nu_1$. 

The facts that  $A_{+1}-A_{-1} <0$ and $\langle\phi_{\pm 1}\rangle-\phi_0 < -\upi/2$ indicate that $3\upi/2 <  2\vartheta_1 -\vartheta_2 < 2\upi$. The phase difference between $\phi_{+2}$ and $\phi_{-2}$, along with this constraint, leads to $2\vartheta_1-\vartheta_2 =5.79\pm 0.05$\,rad. With use of the derived value of $e$, this gives $\varpi=5.85 \pm 0.05$\,rad.

\begin{figure}
\begin{center}
	\includegraphics[width=0.9\linewidth, angle=0]{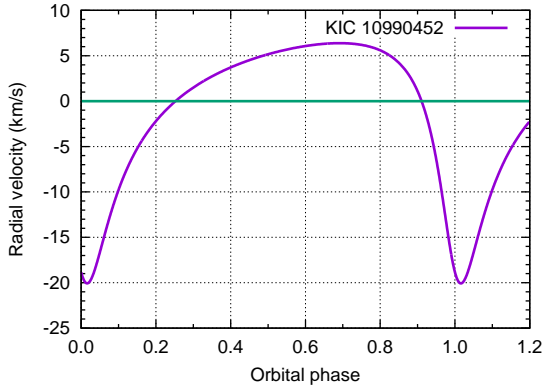} 
\end{center}
\caption{
The radial velocity curve, derived from the Kepler photometric data with the FM method, of KIC\,10990452. The zero point for the orbital phase corresponds to the periapsis passage of the star.} 
\label{fig:11}
\end{figure}

Fig.\,\ref{fig:11} shows the radial velocity curve of KIC\,10990452 derived from the {\it Kepler} photometric data alone. This demonstrates that the present FM method extracts binary information from the {\it Kepler} light curve without spectroscopic observations. 

\begin{table}
\centering
\caption[]{The estimated quantities of the terms for KIC\,10990452, $m\alpha\xi_m$ and $\theta_m$, appearing in the expression of the radial velocity.
They are estimated from the frequency nonuplets for the two highest amplitude modes, $\nu_1$ and $\nu_2$, shown in Table\,\ref{table:04}. 
Column 1 shows the frequency of the central peak, and Column 2 shows its ratio to the orbital frequency. Here $c$ is the speed of light.
Column 4 shows $m\alpha\xi_m$ estimated from equation (\ref{eq:43}). 
Column 5 shows $\theta_m$ estimated by equation (\ref{eq:45}). 
}
\small
\begin{tabular}{ccccr}
\toprule
\multicolumn{1}{c}{$\nu_{\rm puls}$} & 
\multicolumn{1}{c}{$(\nu_{\rm orb}/\nu_{\rm puls})c$} & 
\multicolumn{1}{c}{$m$} & 
\multicolumn{1}{c}{$m\alpha\xi_m$} & 
\multicolumn{1}{c}{$\theta_m-m\theta_1$} \\ 
\multicolumn{1}{c}{d$^{-1}$} & 
\multicolumn{1}{c}{km\,s$^{-1}$} &   
 & 
 & 
\multicolumn{1}{c}{rad}\\ 
\midrule
$17.7237371$ & $138.5 \pm 0.4$ & $1$ & $0.064 \pm 0.001$ & $ 0.000 \pm 0.012$ \\
                        &                            & $2$ & $0.034 \pm 0.002$ & $ 3.632 \pm 0.045$ \\ 
                        &                            & $3$ & $0.025 \pm 0.003$ & $ 4.084 \pm 0.056$ \\ 
                        &                            & $4$ & $0.020 \pm 0.004$ & $ 4.574 \pm 0.156$ \\  
\midrule
$17.8568332$ & $137.4 \pm 0.4$ & $1$ & $0.067 \pm 0.002$ & $ 0.006 \pm 0.024$ \\
                        &                            & $2$ & $0.042 \pm 0.004$ & $ 3.573 \pm 0.077$ \\ 
                        &                            & $3$ & $0.021 \pm 0.006$ & $ 4.263 \pm 0.162$ \\ 
                        &                            & $4$ & $0.008 \pm 0.008$ & $ 4.767 \pm 0.347$ \\ 
\bottomrule
\end{tabular}
\label{table:05}
\end{table}

\begin{table}
\centering
\caption[]{The binary parameters of KIC\,10990452 derived with the FM method.}
\small
\begin{tabular}{cr@{\,$\pm$\,}lr@{\,$\pm$\,}l}
\toprule
\multicolumn{1}{c}{Quantity}& 
\multicolumn{2}{c}{PM} & 
\multicolumn{2}{c}{FM} \\
\midrule
$P_{\rm orb}$ (d) & $122.10$ & $0.21$ & $122.11$ &  $0.36$  \\
$e$                       & $0.55$ & $0.03$ & $0.57$     & $0.03$  \\ 
$\varpi$  (rad)       & $5.81$ & $0.05$ & $5.85$     & $0.05$   \\
$a_1\sin i$  (au)        & $0.123$ & $0.016$ & $0.122$     & $0.023$  \\
\bottomrule
\end{tabular}
\label{table:06}
\end{table}

\begin{figure*}
\centering
\includegraphics[width=0.80\linewidth,angle=0]{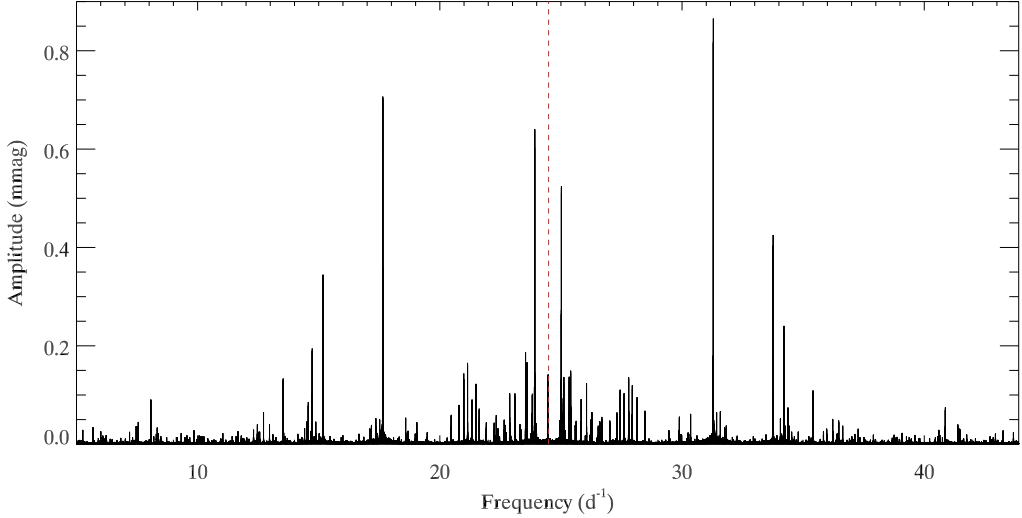}
\caption{Fourier transform for KIC\,8264492 from 5.0 to 43.9\,d$^{-1}$. The highest peak is real and is used in our FM analysis. The frequencies used for the PM analysis are $\nu_1 \dots \nu_9 =$ 31.29, 23.92, 33.76, 34.21, 23.54, 21.14, 24.47, 20.99 and 13.53\,d$^{-1}$. The dashed red line represents the Nyquist frequency of the Kepler LC data.}
\label{fig:12}
\end{figure*}

\begin{table*}
\begin{center}
\caption{A least-squares fit of the frequency undecuplet for the three highest amplitude modes in the KIC\,8264492 data set. The zero-point in time was set at BJD = 2\,455\,754.1266, so that the first sidelobes of the highest amplitude mode had exactly equal phases. From the first pair of sidelobes, we obtain a frequency splitting of $0.0039621\pm0.0000087$\,d$^{-1}$, giving $P_{\rm orb} = 252.39\pm0.56$\,d. Frequency uncertainties were calculated from a separate non-linear least-squares calculation where sidelobes were not forced to be equally split by the orbital frequency, and should therefore be taken as representative, only. Since only three modes were investigated, there is substantial variance left in the data. Treating the remaining variance would reduce the errors. The fact that the differences between the phases of the first sidelobes and the central peak for each mode are so close to $\upi/2$ confirms the binarity within the FM theory.
}
\begin{tabular}{cccrrrrr}
\toprule
$m$ & Frequency & Amplitude & 
\multicolumn{1}{c}{Phase} &
\multicolumn{1}{c}{${\langle\phi_{\pm 1}\rangle}-{\phi_0}$} & 
\multicolumn{1}{c}{  $\phi_{+m}-\phi_{-m}$ } &
\multicolumn{1}{c}{$\frac{A_{+m}+A_{-m}}{A_0}$} &
\multicolumn{1}{c}{$\frac{A_{+m}-A_{-m}}{A_{+m}+A_{-m}}$}\\ 
 & d$^{-1}$ & mmag & 
\multicolumn{1}{c}{rad} & 
\multicolumn{1}{c}{rad} & 
\multicolumn{1}{c}{rad} &
  & \\ 
\midrule
$    -5    $&$     31.2722205      \pm     0.0001598       $&$     0.009   \pm     0.003   $&$     1.495   \pm     0.380   $ &$ $&$ $ & \\
$    -4    $&$     31.2761829      \pm     0.0001108       $&$     0.013   \pm     0.003   $&$     1.065   \pm     0.263   $ &$ $&$ $ & \\
$    -3    $&$     31.2801453      \pm     0.0000539       $&$     0.027   \pm     0.003   $&$     0.244   \pm     0.128   $ &$ $&$ $ & \\
$    -2    $&$     31.2841076      \pm     0.0000300       $&$     0.048   \pm     0.003   $&$     -0.115  \pm     0.071   $ &$ $&$ $ & \\
$    -1    $&$     31.2880700      \pm     0.0000086       $&$     0.167   \pm     0.003   $&$     -0.239  \pm     0.020   $ &$ $&$ $ & \\
$   \phantom{-}0  $&$     31.2920323      \pm     0.0000016       $&$     0.893   \pm     0.003   $&$     1.336   \pm 0.004 $ &$-1.575 \pm 0.015$ & &  \\
$    +1    $&$     31.2959947      \pm     0.0000085       $&$     0.168   \pm     0.003   $&$     -0.239  \pm     0.020   $ & &$0.000 \pm 0.028$ & $0.375 \pm 0.005$ & $0.003 \pm 0.013$\\
$    +2    $&$     31.2999571      \pm     0.0000229       $&$     0.063   \pm     0.003   $&$     -1.083  \pm     0.054   $ & &$-0.968 \pm 0.089$ & $0.124 \pm 0.005$ & $0.135 \pm 0.039$\\
$    +3    $&$     31.3039194      \pm     0.0000330       $&$     0.043   \pm     0.003   $&$     -2.138  \pm     0.078   $ & &$-2.382 \pm 0.081$ &$0.078 \pm 0.005$ & $0.229 \pm 0.062$\\
$    +4    $&$     31.3078818      \pm     0.0000984       $&$     0.015   \pm     0.003   $&$     -2.579  \pm     0.232   $ & &$-3.644 \pm 0.232 $& $0.031 \pm 0.005$ & $ 0.071 \pm 0.152$\\
$    +5    $&$     31.3118441      \pm     0.0001587       $&$     0.009   \pm     0.003   $&$     -2.401  \pm     0.375   $ & &$-3.896 \pm 0.376 $& $0.020 \pm 0.005$ & $0.000 \pm 0.236$\\
\midrule
$     -2    $&$     23.9104185      \pm     0.0000399       $&$     0.036   \pm     0.003   $&$     -2.877  \pm     0.094   $ &$ $&$ $ &\\
$     -1    $&$     23.9143809      \pm     0.0000144       $&$     0.099   \pm     0.003   $&$     2.654   \pm     0.034   $ &$ $&$ $ &\\
$   \phantom{-}0  $&$     23.9183432      \pm     0.0000022       $&$     0.656   \pm     0.003   $&$     -2.102  \pm 0.005 $ &$-1.624 \pm 0.025$& &  \\
$     +1    $&$     23.9223056      \pm     0.0000150       $&$     0.095   \pm     0.003   $&$     2.461   \pm     0.036   $ & &$ -0.193 \pm 0.050$ & $0.296 \pm 0.007$ & $-0.021 \pm 0.022$ \\
$     +2    $&$     23.9262680      \pm     0.0000441       $&$     0.032   \pm     0.003   $&$     1.861   \pm     0.105   $ & &$ 4.738 \pm 0.141$ & $0.104 \pm 0.006$ & $-0.059 \pm 0.062$ \\
\midrule
$      -4    $&$     33.7448167      \pm     0.0001495       $&$     0.010   \pm     0.003   $&$     -2.397  \pm     0.353   $ &$ $&$ $ &\\
$      -3    $&$     33.7487790      \pm     0.0001168       $&$     0.012   \pm     0.003   $&$     -2.600  \pm     0.275   $ &$ $&$ $ &\\
$      -2    $&$     33.7527414      \pm     0.0000533       $&$     0.027   \pm     0.003   $&$     -2.810  \pm     0.125   $ &$ $&$ $ &\\\
$      -1    $&$     33.7567037      \pm     0.0000156       $&$     0.092   \pm     0.003   $&$     -2.975  \pm     0.037   $ &$ $&$ $ &\\
$   \phantom{-}0  $&$     33.7606661      \pm     0.0000034       $&$     0.426   \pm     0.003   $&$     -1.401  \pm 0.008 $ &$-1.625 \pm 0.028 $& &   \\
$      +1    $&$     33.7646285      \pm     0.0000163       $&$     0.088   \pm     0.003   $&$     -3.076  \pm     0.039   $ & & $-0.101 \pm 0.054 $ & $0.423 \pm 0.010$ & $-0.022 \pm 0.024$\\
$      +2    $&$     33.7685908      \pm     0.0000468       $&$     0.031   \pm     0.003   $&$     2.520   \pm     0.110   $ & & $5.330 \pm 0.167 $ & $0.136 \pm 0.010$ & $0.069 \pm 0.073$ \\
$      +3    $&$     33.7725532      \pm     0.0001162       $&$     0.012   \pm     0.003   $&$     1.799   \pm     0.277   $ & & $4.399 \pm 0.280 $ & $0.056 \pm 0.010$ & $0.000 \pm 0.177$ \\
$      +4    $&$     33.7765155      \pm     0.0002028       $&$     0.007   \pm     0.003   $&$     0.969   \pm     0.479   $ & & $3.396 \pm 0.479 $ & $0.040 \pm 0.010$ & $-0.176 \pm 0.253$\\
\bottomrule
\end{tabular}
\label{table:07}
\end{center}
\end{table*}

\subsection{The highly eccentric star KIC\,8264492}
\label{sec:5.3}
\subsubsection{PM analysis}
\label{sec:5.3.1}
KIC\,8264492 is a binary with a $\delta$\,Sct star component. \citet{Huber2014} give $T_{\rm eff} = 7992^{+231}_{-315}$\,K, $\log g = 3.947^{+0.199}_{-0.224}$ (cgs units), and $M = 1.87$\,M$_{\odot}$. The binary nature of KIC\,8264492 was discovered using the PM method \citep{PM2014}. The Fourier transform of its light curve (Fig.\,\ref{fig:12}a) shows a dense frequency spectrum, in which the highest amplitude peak occurs at 31.29\,d$^{-1}$. This peak is distinguished from its Nyquist alias at 17.7\,d$^{-1}$ by the larger amplitude of the former \citep{sNa2013}. The non-sinusoidal nature of the time delays indicated the orbit is highly eccentric, which is confirmed by the high amplitude of the harmonics 
(see figure 21 of \citealt{PM2}). 
From the amplitude ratios of those harmonics, a graphic solution for the eccentricity, $e \simeq 0.73 \pm 0.05$, is obtained with the help of Fig.\,\ref{fig:05}.

\subsubsection{FM analysis}
\label{sec:5.3.2}
The following FM analysis is based on the highest amplitude peak from Fig.\,\ref{fig:12}a. We removed just over 5000 points from the start of the Q0--Q16 LC data set because of some trends and flux discontinuities in the early quarters. The data set we used has 64\,031 LC points, and is 1318.9-d long (BJD 2\,455\,072.035 to 2\,456\,390.959). 

\begin{figure*}
\begin{center}
	\includegraphics[width=0.48\linewidth]{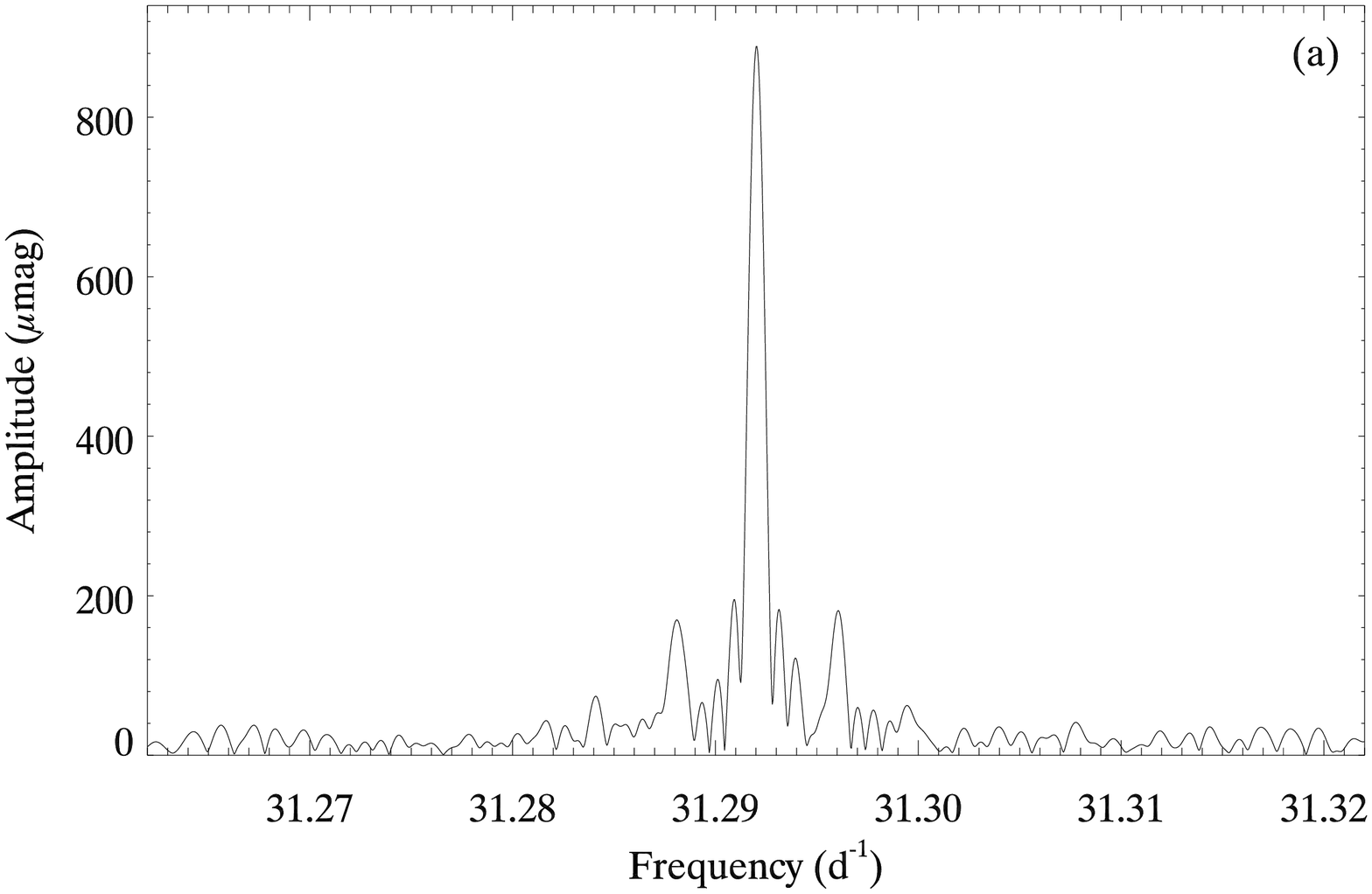} 
	\includegraphics[width=0.48\linewidth]{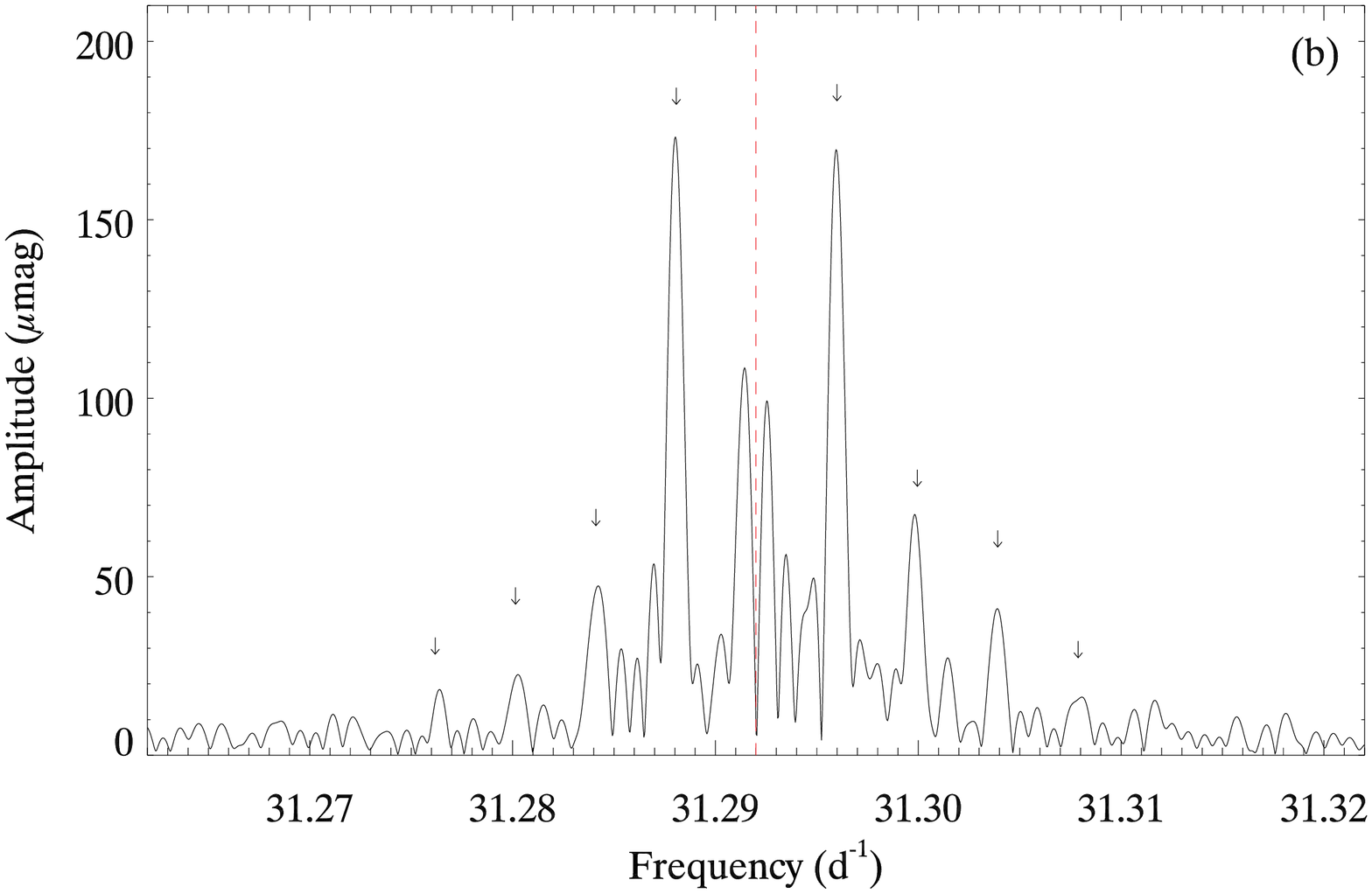} 
\end{center}
\caption{(a): The amplitude spectrum for KIC\,8264492, centred and zoomed on the highest peak. Sidelobes are visible without prewhitening the central peak, but in panel (b) the same region is shown with the central peak prewhitened (at the dashed red line). Arrows are drawn at exact multiples of the orbital frequency distant from the central peak, showing that four pairs of sidelobes are easily identified. 
}
\label{fig:13}
\end{figure*}

Fig.\,\ref{fig:13}a zooms in on the highest peak at \mbox{$\nu_1=31.2920323\,{\rm d}^{-1}$}. The first sidelobes are clearly visible, separated from the central peak by $0.004$\,d$^{-1}$. Fig.\,\ref{fig:13}b shows the amplitude spectrum after the central peak (at the dashed red line) has been prewhitened. 
The FM sidelobes are apparent, as shown by arrows, at exact multiples of the orbital frequency distant from the central peak.

Table~\ref{table:07} gives the frequencies, amplitudes, phases and their uncertainties for the three highest amplitude modes and their detectable pairs of sidelobes. We determined the orbital frequency splitting from the splitting of the first pair of sidelobes with respect to the central component, $\nu_{\rm orb} = 0.0039621\pm0.0000087$\,d$^{-1}$, giving $P_{\rm orb} = 252.39\pm0.56$\,d. 

For each set of sidelobes, the values of $m\alpha\xi_m$ and $\theta_m-m\theta_1$ were derived by equations (\ref{eq:43}) and (\ref{eq:45}), respectively, and are summarized in Table\,\ref{table:08}. The eccentricity $e$ is then estimated from the first and second sidelobes around $\nu_1$ by equation (\ref{eq:04new4}) to be $e = 0.761 \pm 0.045$.

The technique explained in Section 4.1 and inspection of Fig.\,\ref{fig:03}  gives the coefficient $\xi_1(e, \varpi) = 0.630\pm 0.030$. Using this value and the amplitude ratio for $\nu_1$, $\alpha\xi_1=0.369\pm 0.004$, we obtain $\alpha = 0.577\pm 0.028$. The mass function is determined from these values to be $f(m_1,m_2,\sin i)=0.274\pm 0.040\,{\rm M}_{\odot}$. Assuming $m_1=1.87$\,M$_\odot$ gives a minimum secondary mass of $m_2 = 1.44\pm 0.10$\,M$_{\odot}$. 
We also derive the semi-major axis of the primary star about the barycentre. From equation (\ref{eq:21}), we find $a_1 \sin i = 0.508 \pm 0.024$\,au from the data for $\nu_1$. 

The minimum mass derived above puts the star in the mid-F range, or more massive. That suggests that some of the peaks seen in Fig.\,\ref{fig:12}a may come from pulsations in the secondary. 
However, no evidence of anti-phase time delay variations in p modes are found, and 
g modes are low amplitude and not PM sensitive.

The facts that  $A_{+2}-A_{-2} >0$ for $\nu_1$ and $\langle\phi_{\pm 1}\rangle-\phi_0 < -\upi/2$ for $\nu_2$ indicate $3\upi/2 <  2\vartheta_1 -\vartheta_2 < 2\upi$.
The phase difference between $\phi_{+2}$ and $\phi_{-2}$, along with this constraint, leads to $2\vartheta_1-\vartheta_2 =5.20\pm 0.04$\,rad. With use of the derived value of $e$, this gives $\varpi=5.28 \pm 0.04$\,rad.

\begin{figure}
\begin{center}
	\includegraphics[width=0.9\linewidth, angle=0]{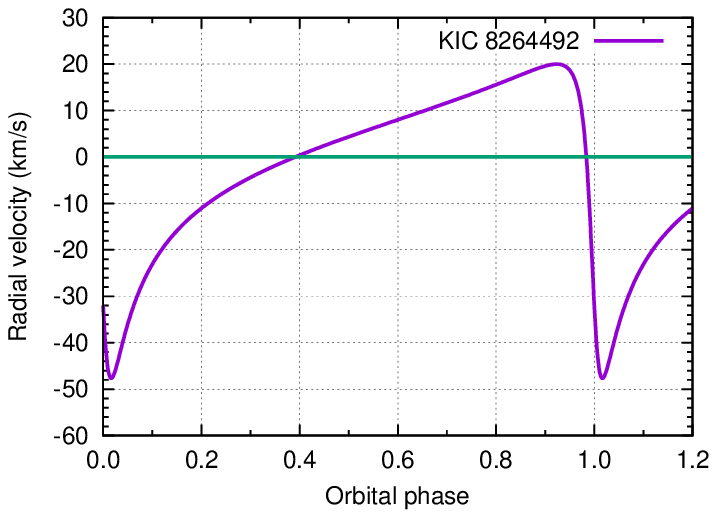} 
	\includegraphics[width=0.9\linewidth, angle=0]{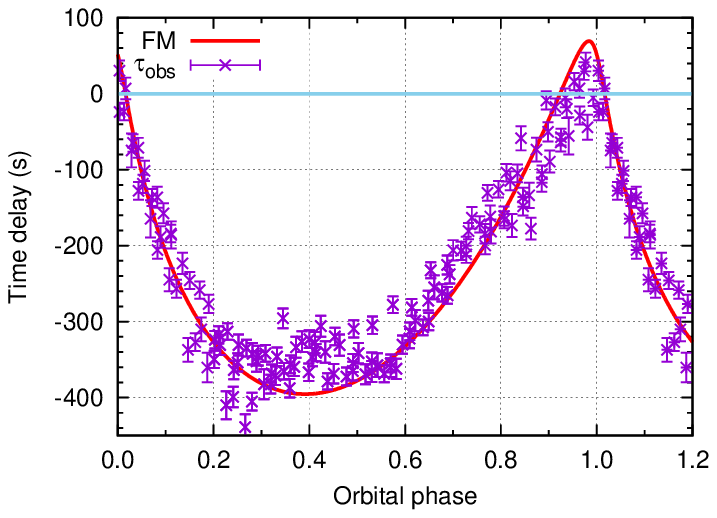} 
\end{center}
\caption{
{\bf Upper panel:} The radial velocity curve, derived from the Kepler photometric data with the FM method, of KIC\,8264492. The zero point for the orbital phase corresponds to the periapsis passage of the star. 
{\bf Lower panel:} The time delay of KIC\,8264492, derived with the FM method. The time delay shown here is the difference in the light arrival time, compared with the case where the star lies at the barycentre of the binary system. 
}
\label{fig:14}
\end{figure}

The upper panel of Fig.\,\ref{fig:14} shows the radial velocity curve of KIC\,8264492 derived with the FM method from the {\it Kepler} photometric data alone. The time delay is also derived with the FM method. The lower panel of Fig.\,\ref{fig:14} shows the variation in the light arrival time with respect to the barycentre. This is in good agreement with time delays derived by the PM method 
(see \cite{PM2}).

\begin{table}
\centering
\caption[]{The estimated quantities of the terms for KIC\,8264492, $m\alpha\xi_m$ and $\theta_m$, appearing in the expression of the radial velocity.
They are estimated from the frequency nonuplets for the three highest amplitude modes, $\nu_1$, $\nu_2$ and $\nu_3$, shown in Table\,\ref{table:07}. 
Column 1 shows the frequency of the central peak, and Column 2 shows its ratio to the orbital frequency. Here $c$ is the speed of light.
Column 4 shows $m\alpha\xi_m$ estimated from equation (\ref{eq:43}). 
Column 5 shows $\theta_m$ estimated by equation (\ref{eq:45}). 
}
\small
\begin{tabular}{ccccr}
\toprule
\multicolumn{1}{c}{$\nu_{\rm puls}$} & 
\multicolumn{1}{c}{$(\nu_{\rm orb}/\nu_{\rm puls})c$} & 
\multicolumn{1}{c}{$m$} & 
\multicolumn{1}{c}{$m\alpha\xi_m$} & 
\multicolumn{1}{c}{$\theta_m-m\theta_1$} \\ 
\multicolumn{1}{c}{d$^{-1}$} & 
\multicolumn{1}{c}{km\,s$^{-1}$} &   
 & 
 & 
\multicolumn{1}{c}{rad}\\ 
\midrule
$31.2920323$ & $38.0 \pm 0.1$ & $1$ & $0.369 \pm 0.004$ & $ 0.000 \pm 0.014$ \\
                        &                          & $2$ & $0.248 \pm 0.010$ & $ 1.087 \pm 0.045$ \\ 
                        &                          & $3$ & $0.234 \pm 0.015$ & $ 1.951 \pm 0.040$ \\ 
                        &                          & $4$ & $0.124 \pm 0.020$ & $ 2.890 \pm 0.116$ \\  
                        &                          & $5$ & $0.100 \pm 0.025$ & $ 4.335 \pm 0.188$ \\  
\midrule
$23.9183432$ & $49.7 \pm 0.1$ & $1$ & $0.293 \pm 0.007$ & $-0.097 \pm 0.025$ \\ 
                        &                          & $2$ & $0.208 \pm 0.012$ & $ 3.940 \pm 0.071$ \\ 
\midrule
$33.7606661$ & $35.2 \pm 0.1$ & $1$ & $0.414 \pm 0.009$ & $-0.051 \pm 0.027$ \\
                        &                          & $2$ & $0.272 \pm 0.020$ & $ 4.236 \pm 0.083$ \\ 
                        &                          & $3$ & $0.168 \pm 0.030$ & $ 5.344 \pm 0.014$ \\ 
                        &                          & $4$ & $0.160 \pm 0.040$ & $ 6.395 \pm 0.240$ \\ 
\bottomrule
\end{tabular}
\label{table:08}
\end{table}

\begin{table}
\centering
\caption[]{Comparison of the binary parameters of KIC\,8264492 derived with the FM method and the PM method.}
\small
\begin{tabular}{cr@{\,$\pm$\,}lr@{\,$\pm$\,}l}
\toprule
\multicolumn{1}{c}{Quantity}& 
\multicolumn{2}{c}{PM} & 
\multicolumn{2}{c}{FM} \\
\midrule
$P_{\rm orb}$ (d) & $253.78$ &  $1.03$  & $252.39$ & $0.56$ \\
$e$                       & $0.67$   & $0.04$   & $0.76$    & $0.05$ \\ 
$\varpi$  (rad)       &  $5.28$  & $0.05$   & $5.28$   & $0.04$ \\
$a_1\sin i$  (au)        &  $0.41$  & $0.05$   & $0.51$   & $0.02$ \\
$f(m_1,m_2,\sin i)$ (M$_\odot$) &  $0.143$  & $0.054$ & $0.274$ & $0.040$ \\
\bottomrule
\end{tabular}
\label{table:09}
\end{table}

\section{Discussion and conclusions}
\label{sec:6}

We have shown in this paper that the Fourier transform of the light curve of a pulsating star in a binary system leads to frequency multiplets in the amplitude spectra 
and that all the binary orbital information traditionally derived from spectroscopic radial velocity measurements can be derived from the frequency splitting and the amplitudes and phases of the components of the frequency multiplet. This is an extension of the FM method described in \citet{FM2012}, in which the case of $e \ll 1$ was mainly discussed.  We have improved the applicability to highly eccentric cases, and presented a method for determining binary orbital parameters, including the eccentricity and the argument of periapsis, from photometry alone.

The orbital elements to be determined are (i) the orbital angular frequency, $\Omega$, (ii) the projected semi-major axis, $a\sin i$, (iii) the eccentricity, $e$, and (iv) the argument of the periapsis, $\varpi$. Hence, we need at least four independent relations among the components of the frequency multiplet. The orbital angular frequency is directly measured from the frequency spacing of the adjacent components of the multiplet. The combination of the amplitude ratio of the sum of a pair of $\pm 1$-components to the central component, $(A_{+1}+A_{-1})/A_0$, and the ratio for the case of $\pm 2$-components, $(A_{+2}+A_{-2})/A_{0}$, gives $a_1\sin i$ and $e$. The phase information of the first and the second sidelobes leads to the argument of the periapsis. Careful examination of the phase differences also indicates whether the periapsis is located at the far side or the near side of the orbit, with respect to us. Hence, even if both components of the binary are pulsating, we can separate the pulsation spectra of two pulsating stars.

Binary information can be extracted by the PM method 
\citep{PM2014, PM2}. 
The present FM method and the PM analysis are complementary. The PM analysis offers a clear visualization of the binary orbit in the time domain by dividing the light curve into segments, while the present FM method utilises the full frequency resolution of the data in the Fourier domain, and leads to high precision, particularly in the case of short-period binaries.

For pulsating stars, such as $\delta$~Sct stars, with many significant peaks in the amplitude spectrum, some care is needed to test for possible unresolved contamination of FM multiplet components by other, independent mode or combination frequencies. This is easily done for multi-periodic pulsators, because the amplitude and phase relationships for each FM multiplet must agree within the relations given in this paper.  In the PM method, unresolved contamination results in obvious outliers in the time delay diagram, again making it easy recognise and discard contaminated mode frequencies.

A promising application of our method is in the search for exoplanets orbiting pulsating $\delta$\,Scuti stars, or more generally, exoplanets around upper main-sequence stars. Both the transit method and the ground-based Doppler method, which are widely used in the search for exoplanets around solar-like stars and have discovered most of the known exoplanets to date, are more difficult to apply for upper main-sequence stars. Planetary transits are smaller in the case of upper main-sequence stars because of the much larger stellar disks. On the other hand, the typical rotationally broadened spectral lines of A-type and earlier type stars make it harder to detect the Doppler shift of spectral lines. Our new technique of photometric measurement of binary orbital parameters opens a possibility of exoplanet detection around upper main sequence pulsating stars, such as $\delta$~Sct stars, and around compact pulsating stars. 

Another application of the present method is at the opposite extreme, that is, in the search for invisible massive companions in binary systems. If a pulsating $\delta$~Sct star forms a binary system with a star born initially with mass less than about $8\,{\rm M}_{\odot}$ but more than the pulsating star itself, the massive component must have already evolved  into a white dwarf while the less massive component is still in the main-sequence stage as a pulsating $\delta$~Sct star.
The mass of the white dwarf must be less than the Chandrasekhar mass limit, but still be of $\sim 1\,{\rm M}_{\odot}$, and the star must be too faint to be detected.
On the other hand, if a binary system is composed of a $\delta$~Sct star and a star born initially more massive than $\sim 8\,{\rm M}_{\odot}$ but less than $\sim 30\,{\rm M}_{\odot}$, the massive star must have already been turned to a neutron star, probably with mass $\sim 1\,{\rm M}_\odot$, and the star must be again undetectable except perhaps as a pulsar. Although the survival of the companion after the explosion of the massive star is uncertain from the theoretical viewpoint, 
we know of binary systems composed of a neutron star and either an early-type massive star, as a high-mass X-ray binary (HMXB), or a cool star, as a low-mass X-ray binary (LMXB). Hence it seems natural to expect a binary system of a neutron star and an A-type $\delta$~Sct star. The further extreme case is a star born with an initial mass $\gtrsim 30\,{\rm M}_{\odot}$. Such a star becomes a black hole. A well-known binary system composed of a black hole candidate is Cygnus X-1, whose companion is a blue supergiant variable star. 
If a black hole forms a binary system with a $\delta$\,Sct star, the present FM technique is a unique, promising method of finding such an exotic system.

\section*{acknowledgements}
This work was initiated with partial support from a Japan-UK Joint Research project of 
the Japan Society for the Promotion of Science (JSPS), and also from
a Royal Society UK-Japan International Joint program. This work was completed during a JSPS invitation fellowship to DWK, for which he thanks JSPS. 
This research was also supported by the Australian Research Council. Funding for the Stellar Astrophysics Centre is provided by the Danish National Research Foundation (grant agreement no.: DNRF106). The research is supported by the ASTERISK project (ASTERoseismic Investigations with SONG and Kepler) funded by the European Research Council (grant agreement no.: 267864).

\bibliography{fmbib}

\section*{Appendix 1: Derivation of Equations (11) and (12)}
\renewcommand{\theequation}{A.\arabic{equation}}
\setcounter{equation}{0}

We need to write the radial velocity as an explicit function of time. In doing so, the essential point is how to express trigonometric functions of the true anomaly $f$ as an explicit function of time. In this Appendix, we derive these expressions. 

\renewcommand{\thesubsection}{A.\arabic{subsection}}
\setcounter{subsection}{0}
\subsection{Derivation of equation (\ref{eq:11})}
Let us first consider $\cos f$. The distance $r$ between the barycentre and the star is expressed with help of a combination of the semi-major axis $a_1$, the eccentricity $e$ and the true anomaly $f$:
\begin{equation}
	r = {{a_1(1-e^2)}\over{1+e\cos f}},
\label{eq:a.1}
\end{equation}
and then
\begin{eqnarray}
\lefteqn{
	\cos f = {{1}\over{e}}\left[ (1-e^2){{a_1}\over{r}}-1\right].
}
\label{eq:a.2}
\end{eqnarray}
Since, with the help of the eccentric anomaly $u$, as derived from equations (\ref{eq:8}) and (\ref{eq:9}),
\begin{equation}
	r = a_1(1-e\cos u),
\label{eq:a.3}
\end{equation}
the term $a_1/r$ in the right-hand-side of equation (\ref{eq:a.2}) is
\begin{equation}
	{{a_1}\over{r}} = {{1}\over{1-e\cos u}}.
\label{eq:a.4}
\end{equation}
Kepler's equation links the eccentric anomaly $u$ with the time after the periapsis passage:
\begin{equation}
	u - e\sin u = \Omega (t-t_{\rm p}).
\label{eq:a.5}
\end{equation}
The partial derivative of $u$ with respect to $t$ leads to
\begin{equation}
	{{\partial u}\over{\partial t}}={{\Omega}\over{1-e\cos u}},
\label{eq:a.6}
\end{equation}
and then, with the help of equations (\ref{eq:a.4}) and (\ref{eq:a.6}), equation (\ref{eq:a.2}) is reduced to 
\begin{equation}
	\cos f = {{1}\over{e}}\left[ (1-e^2){{1}\over{\Omega}}{{\partial u}\over{\partial t}}-1\right].
\label{eq:a.7}
\end{equation}
The eccentric anomaly $u$ is expressed as a solution of Kepler's equation, in terms of Fourier expansion with respect to the time after periapsis passage
[see textbooks on celestial mechanics, e.g., \cite{brouwerclemence1961}]: 
\begin{equation}
	u=\Omega (t-t_{\rm p}) + 2\sum_{n=1}^\infty {{1}\over{n}} J_n(ne)\sin n\Omega (t-t_{\rm p}).
\label{eq:a.8}
\end{equation}
Then, 
\begin{equation}
	{{1}\over{\Omega}}{{\partial u}\over{\partial t}}
	=
	1 + 2\sum_{n=1}^\infty J_n(ne)\cos n\Omega(t-t_{\rm p}).
\label{eq:a.9}
\end{equation}
Substituting equation (\ref{eq:a.9}) into equation (\ref{eq:a.7}), we reach the expression given in equation (\ref{eq:11}): 
\begin{equation}
	\cos f
	= -e + {{2(1-e^2)}\over{e}}\sum_{n=1}^\infty J_n(ne)\cos n\Omega (t-t_{\rm p}).
\label{eq:a.10}
\end{equation}

\subsection{Derivation of equation (\ref{eq:12})}
Next, let us consider $\sin f$.
From equation (\ref{eq:9})
\begin{equation}
	\sin f = \sqrt{1-e^2} {{a_1}\over{r}}\sin u. 
\label{eq:a.11}
\end{equation}
The partial derivative of $u$ given in equation (\ref{eq:a.5}) with respect to $e$ leads to
\begin{equation}
	{{\partial u}\over{\partial e}} = {{\sin u}\over{1-e\cos u}} .
\label{eq:a.12}
\end{equation}
Then, with the help of equation (\ref{eq:a.4}), 
\begin{equation}
	\sin f = \sqrt{1-e^2} {{\partial u}\over{\partial e}} .
\label{eq:a.13}
\end{equation}
On the other hand, from equation (\ref{eq:a.8}),
\begin{equation}
	{{\partial u}\over{\partial e}}
	=
	2\sum_{n=1}^\infty J_n{'}(ne) \sin n\Omega(t-t_{\rm p}),
\label{eq:a.14}
\end{equation}
where $J_n{'}(x) := {\rm d}J_n(x)/{\rm d}x$. Hence, we obtain the expression given in equation (\ref{eq:12}):
\begin{equation}
	\sin f = 2\sqrt{1-e^2} \sum_{n=1}^\infty J_n{'}(ne) \sin n\Omega (t-t_{\rm p}).
\label{eq:a.15}
\end{equation}

\section*{Appendix 2: A Frequency Undecuplet}
In Section\,\ref{sec:3.2}, we truncated the infinite series of equation (\ref{eq:27}) with $N=5$,
and derived the relations between the complex amplitudes of the first and second sidelobes.
The complex amplitudes of the higher order sidelobes of the undecuplet are given as follows.
\begin{eqnarray}
\lefteqn{
	{{{\cal A}_{\pm 1}}\over{{\cal A}_0}}
	=
	\pm \left( {{ J_{1}(\alpha\xi_1) }\over{ J_0(\alpha\xi_1) }} \right)	
	\left[ 
	\left\{
	1 
	\mp 
	\left( {{ J_{1}(\alpha\xi_2) } \over {J_0(\alpha\xi_2)  }} \right)
	\cos\left(2\vartheta_1 - \vartheta_2\right)
	\right\}
	\right.
}
	\nonumber\\
\lefteqn{
	\quad\quad\qquad
	\left.
	+ \,{\rm i} \left( {{J_1(\alpha\xi_2)}\over{J_0(\alpha\xi_2)}} \right)
	 \sin(2\vartheta_1-\vartheta_2) 
	 \right]
	\,{\rm e}^{\pm {\rm i}\theta_1},	
}
\label{eq:a.16}
\end{eqnarray}

\begin{eqnarray}
\lefteqn{
	{{{\cal A}_{\pm 2}}\over{{\cal A}_0}}
	\simeq
	\pm \left( {{ J_{1}(\alpha\xi_2) }\over{ J_0(\alpha\xi_2) }} \right)	
}
	\nonumber\\
\lefteqn{
	\quad
	\times
	\left\{ 
	\left[ 
	1 \pm	
	\left( {{J_{1}(\alpha\xi_2)}\over{J_{0}(\alpha\xi_2)}} \right)^{-1}
	\left( {{ J_{2}(\alpha\xi_1) } \over {J_0(\alpha\xi_1)  }} \right)%
	\cos \left( 2\vartheta_1 - \vartheta_2 \right)
	\right]
	\right.
}
	\nonumber\\
\lefteqn{
	\quad
	+ \,{\rm i} 
	\left.
	\left[
	\left( {{J_{1}(\alpha\xi_2)}\over{J_{0}(\alpha\xi_2)}} \right)^{-1}
	\left( {{J_2(\alpha\xi_1)}\over{J_0(\alpha\xi_1)}} \right)
	\sin(2\vartheta_1-\vartheta_2) 
	\right]
	\right\}
	\,{\rm e}^{\pm {\rm i} \theta_2}.	
}
\label{eq:a.17}
\end{eqnarray}

\begin{eqnarray}
\lefteqn{
	{{{\cal A}_{\pm 3}}\over{{\cal A}_0}}
	=
	\pm \left( {{ J_{1}(\alpha\xi_3) }\over{ J_0(\alpha\xi_3) }} \right)	
	\left\{\left[ 
	1 
	\pm 
	\left( {{ J_{1}(\alpha\xi_3) } \over {J_0(\alpha\xi_3)  }} \right)^{-1}
	\right.\right.
}
	\nonumber\\
\lefteqn{
	\quad\quad\qquad
	\left.
	\times\left( {{J_{1}(\alpha\xi_1)J_{1}(\alpha\xi_2)}
	\over{J_{0}(\alpha\xi_1)J_{0}(\alpha\xi_2)}} \right)
	\cos\left(\vartheta_1 + \vartheta_2 - \vartheta_3\right)
	\right] 
}
	\nonumber\\
\lefteqn{
	\quad\quad\qquad
	- \,{\rm i} 
	\left[
	\left( {{ J_{1}(\alpha\xi_3) } \over {J_0(\alpha\xi_3)  }} \right)^{-1}
	\left( {{J_{1}(\alpha\xi_1)J_{1}(\alpha\xi_2)}
	\over{J_{0}(\alpha\xi_1)J_{0}(\alpha\xi_2)}} \right)
	\right.
}
	\nonumber\\
\lefteqn{
	\quad\quad\qquad\qquad
	\times\sin(\vartheta_1+\vartheta_2-\vartheta_3) 
	\biggr]\biggr\}
	\,{\rm e}^{\pm {\rm i}\theta_3} ,	
}
\label{eq:a.18}
\end{eqnarray}

\begin{eqnarray}
\lefteqn{
	{{{\cal A}_{\pm 4}}\over{A_{0}}} 
	\simeq 
	\pm
	\left( {{J_{1}(\alpha\xi_4)}\over{J_{0}(\alpha\xi_4)}} \right) 
	\left\{\left[
	1 \pm
	\left( {{J_{1}(\alpha\xi_4)}\over{J_{0}(\alpha\xi_4)}} \right)^{-1} 
	\right. \right.
}
	\nonumber\\
\lefteqn{
	\quad\quad\qquad
	\times
	\left\{
	\left( {{J_{2}(\alpha\xi_2)}\over{J_{0}(\alpha\xi_2)}} \right)
	\cos(2\vartheta_2-\vartheta_4)
	\right.
}
	\nonumber\\
\lefteqn{
	\quad\quad\quad\qquad
	\left. \left.
	+
	\left(
	{{J_{1}(\alpha\xi_1)J_{1}(\alpha\xi_3)}
	\over{J_{0}(\alpha\xi_1)J_{0}(\alpha\xi_3)}} 
	\right) 
	\cos(\vartheta_1+\vartheta_3-\vartheta_4)
	\right\} \right]
}
	\nonumber\\
\lefteqn{
	\quad\quad\qquad
	+\,{\rm i}
	\left[
	\left( {{J_{1}(\alpha\xi_4)}\over{J_{0}(\alpha\xi_4)}} \right)^{-1} 
	\left( {{J_{2}(\alpha\xi_2)}\over{J_{0}(\alpha\xi_2)}} \right)
	\sin(2\vartheta_2-\vartheta_4)
	\right.
}
	\nonumber\\
\lefteqn{
	\quad\quad\quad\qquad
	+
	\left( {{J_{1}(\alpha\xi_4)}\over{J_{0}(\alpha\xi_4)}} \right)^{-1} 
	\left(
	{{J_{1}(\alpha\xi_1)J_{1}(\alpha\xi_3)}
	\over{J_{0}(\alpha\xi_1)J_{0}(\alpha\xi_3)}} 
	\right) 
}
	\nonumber\\
\lefteqn{
	\quad\quad\qquad\qquad
	\times
	\sin(\vartheta_1+\vartheta_3-\vartheta_4)
	\biggr] \biggr\}
	\,{\rm e}^{\pm {\rm i} \theta_4},
}
\label{eq:a.19}
\end{eqnarray}
and
\begin{eqnarray}
\lefteqn{
	{{{\cal A}_{\pm 5}}\over{{\cal A}_{0}}} = 
	\pm \left(
	{{J_{1}(\alpha\xi_5)}\over{J_{0}(\alpha\xi_5)}} \right)
	\left\{ \left[
	1\pm
	\left( {{J_{1}(\alpha\xi_5)}\over{J_{0}(\alpha\xi_5)}} \right)^{-1} 
	\right. \right.
}
	\nonumber\\
\lefteqn{
	\quad\quad\quad\qquad
	\times
	\left\{
	\left( {{J_{1}(\alpha\xi_2)J_{1}(\alpha\xi_3)}
	\over{J_{0}(\alpha\xi_2)J_{0}(\alpha\xi_3)}} \right)
	\cos(\vartheta_2 + \vartheta_3 - \vartheta_5)
	\right.
}
	\nonumber\\
\lefteqn{
	\quad\quad\quad\qquad
	\left.\left.
	+
	\left({{J_{1}(\alpha\xi_1)J_{1}(\alpha\xi_4)}
	\over{J_{0}(\alpha\xi_1)J_{0}(\alpha\xi_4)}} \right) 
	\cos(\vartheta_1 + \vartheta_4 - \vartheta_5)
	\right\} \right]
}
	\nonumber\\
\lefteqn{
	\quad\quad\qquad
	+ \,{\rm i}
	\left( {{J_{1}(\alpha\xi_5)}\over{J_{0}(\alpha\xi_5)}} \right)^{-1} 
}
	\nonumber\\
\lefteqn{
	\quad\quad\quad\qquad
	\times
	\left[
	\left( {{J_{1}(\alpha\xi_2)J_{1}(\alpha\xi_3)}
	\over{J_{0}(\alpha\xi_2)J_{0}(\alpha\xi_3)}} \right)
	\sin(\vartheta_2 + \vartheta_3 - \vartheta_5)
	\right.
}
	\nonumber\\
\lefteqn{
	\quad\quad\qquad\qquad
	+
	\left({{J_{1}(\alpha\xi_1)J_{1}(\alpha\xi_4)}
	\over{J_{0}(\alpha\xi_1)J_{0}(\alpha\xi_4)}} \right) 
}
	\nonumber\\
\lefteqn{
	\quad\qquad\qquad\qquad
	\times
	\sin(\vartheta_1 + \vartheta_4 - \vartheta_5)
	 \biggr] \biggr\}
	 \,{\rm e}^{\pm {\rm i} \theta_5}.
}
\label{eq:a.20}
\end{eqnarray}

\end{document}